\documentclass[sigconf, nonacm]{acmart}
\usepackage{tabularx}
\usepackage{booktabs}

\usepackage{fancyhdr}
\usepackage[linesnumbered,vlined,ruled]{algorithm2e}
\usepackage{algpseudocode}
\usepackage{graphicx}
\usepackage{textcomp}
\usepackage{xcolor}
\usepackage{bm}
\usepackage{booktabs}
\usepackage{multirow}
\usepackage{enumitem}
\usepackage{microtype}
\usepackage{booktabs} 
\usepackage{amsfonts} 

\usepackage{amssymb}
\usepackage{amsmath}
\usepackage{amsthm}
\usepackage{sansmath}
\usepackage{subfigure}
\usepackage{graphicx}
\usepackage{subcaption}
\usepackage{amsthm}
\usepackage{tcolorbox}
\usepackage{colortbl}
\usepackage{tikz}
\usepackage{url}
\usepackage{pifont}
\usepackage{makecell}
\usepackage{cleveref}
\usepackage{pgfplots}
\usepackage{setspace}
\usepackage{xspace}
\usepackage[dvipsnames]{xcolor}

\usepackage{makecell}

\usepackage{enumerate}

\usepackage{caption}
\setcopyright{none}
\renewcommand\footnotetextcopyrightpermission[1]{}

\definecolor{LightCyan}{rgb}{0.88,1,1}
\definecolor{LightRed}{rgb}{1,0.88,1}
\definecolor{LightYellow}{rgb}{1,1,0.88}
\definecolor{LightGray}{gray}{0.8}

\crefformat{section}{\S#2#1#3} 
\crefformat{subsection}{\S#2#1#3}
\long\def\comment#1{}

\AtBeginDocument{%
  }

\newcommand{\nthesection}{\arabic{section}}

\newcounter{remark}[section]
\renewcommand{\theremark}{\nthesection.\arabic{remark}}

\newcommand{\proofsketch}{\noindent{\bf Proof Sketch: }}

\newcommand{\eop}{\hspace*{\fill}\mbox{$\Box$}\vspace*{1ex}}
\newcommand{\stitle}[1]{\vspace{1ex} \noindent{\bf #1}}

\newcommand{\kw}[1]{{\ensuremath {\mathsf{#1}}}\xspace}

\newcommand{\beqn}{\begin{eqnarray*}}
\newcommand{\eeqn}{\end{eqnarray*}}

\newcounter{ccc}

\newcommand{\AGNews}{{\kw{AGNews}}\xspace}
\newcommand{\GSM}{{\kw{GSM8K}}\xspace}
\newcommand{\IMDB}{{\kw{IMDB}}\xspace}
\newcommand{\MMLU}{{\kw{MMLU}}\xspace}
\newcommand{\MRPC}{{\kw{MRPC}}\xspace}
\newcommand{\SNLI}{{\kw{SNLI}}\xspace}

\newcommand{\RoBatch}{{\textsc{RoBatch}}\xspace}

\newcommand{\RouteLLM}{{\text{RouteLLM}}\xspace}
\newcommand{\FrugalGPT}{\text{FrugalGPT}\xspace}
\newcommand{\BATCHER}{{\text{BATCHER}}\xspace}
\newcommand{\BATCHERSIM}{{\text{BATCHER}\mbox{-}{SIM}}\xspace}
\newcommand{\BATCHERDIV}{{\text{BATCHER}\mbox{-}{DIV}}\xspace}
\newcommand{\OBP}{{\text{OBP}}\xspace}

\newcounter{example}

\begin{document}

\title{Towards Cost-effective LLMs Routing with Batch Prompting}

\author{Haotian Xu}
\email{haotianxu@bit.edu.cn}
\affiliation{%
  \institution{Beijing Institute of Technology}
  \city{Beijing}
  \country{China}
}

\author{Kangfei Zhao}
\authornote{Corresponding author.}
\email{zkf1105@gmail.com}
\affiliation{%
  \institution{Beijing Institute of Technology}
    \city{Beijing}
  \country{China}
}

\author{Jiadong Xie}
\email{jdxie@se.cuhk.edu.hk}
\affiliation{%
  \institution{The Chinese University of Hong Kong}
    \city{Hong Kong SAR}
  \country{China}
}

\begin{abstract}

Large Language Model (LLM) serving systems must balance task performance against monetary cost. Two prominent optimization techniques have emerged independently: LLM routing, which directs each query to the most cost-effective model in a model pool, and batch prompting, which packs multiple queries into a single invocation to amortize the fixed cost of the shared system prompt. These two techniques are logically complementary; i.e., routing optimizes the model assignment dimension while batching optimizes the query aggregation dimension, jointly reshaping the landscape of model utility and monetary cost. However, existing approaches explore only one side of this decision space. 
On the basis of empirical studies on their impacts, we are motivated to jointly optimize these two dimensions in this paper. 

We formulate the \emph{Route with Batching Problem}, which jointly determines the target model and batch size for each query under a total cost budget, and prove it NP-hard. 
To solve this challenging problem, we propose \RoBatch, a unified two-stage framework. In the modeling stage, \RoBatch constructs a batch-aware proxy utility model that decomposes combinatorial utility estimation into utility estimation without batching and recalibration of model-specific utility degradation with batching. 
In the routing stage, \RoBatch employs a greedy scheduling algorithm that progressively upgrades the assignment of the target model and batch size for queries along the cost-utility Pareto frontier until the budget is exhausted.
Extensive experiments on six benchmarks across two LLM families (Qwen3 and Gemma3) demonstrate that \RoBatch consistently achieves a superior cost–performance Pareto frontier compared with LLM routing and batch prompting baselines.

\end{abstract}

\maketitle

\section{Introduction}

Large Language Models (LLMs) are rapidly becoming a foundational component of modern data- and AI-intensive applications. 
They power a wide spectrum of applications, including question answering~\cite{DBLP:journals/pvldb/ZhuCXLSZSTL24}, content understanding~\cite{DBLP:conf/naacl/DevlinCLT19,DBLP:conf/sigmod/LiWZW25}, classification~\cite{DBLP:journals/corr/abs-1907-11692}, reasoning~\cite{DBLP:conf/nips/Wei0SBIXCLZ22}, recommendation~\cite{DBLP:journals/tkde/ZhaoFLLMWWWZTL24, qi2026sema}, code assistance~\cite{DBLP:conf/emnlp/0034WJH21}, and increasingly, agentic workflows~\cite{DBLP:conf/iclr/YaoZYDSN023,DBLP:journals/pvldb/ShankarCSPW25,DBLP:journals/pvldb/ChangG25} that invoke models as part of data-processing pipelines. 
As LLMs move from standalone demos to production-level services, the focus is no longer merely on how to obtain a good answer from a single model invocation, but how to serve large volumes of query instances economically and reliably. This means that LLM serving systems must balance task serving quality, monetary expenditure, and response efficiency, etc. under practical workloads and budget constraints. 

\begin{figure*}[ht]
     \begin{tabular}[h]{c}
         \subfigure[LLM Routing] {
		      \includegraphics[width=0.64\columnwidth]{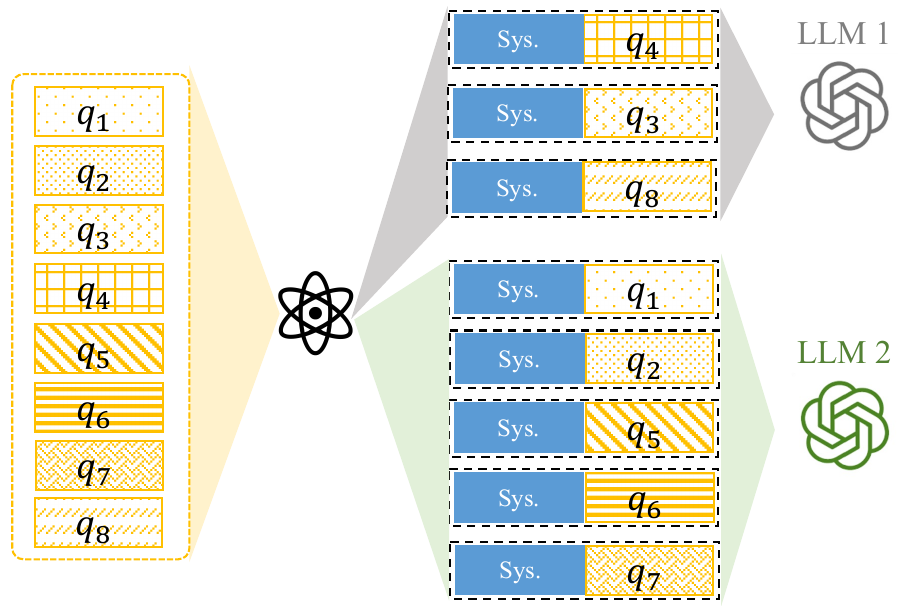}
			\label{fig:intro:route}
		} 
        
        \subfigure[Batch Prompting] {
		      \includegraphics[width=0.64\columnwidth]{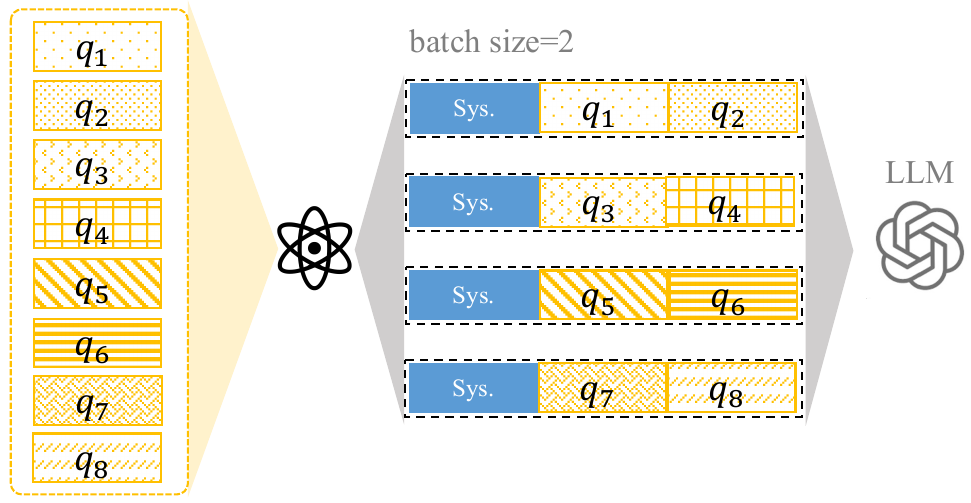}
			\label{fig:intro:batch}
		} 
        \subfigure[Route with Batching] {
			\includegraphics[width=0.64\columnwidth]{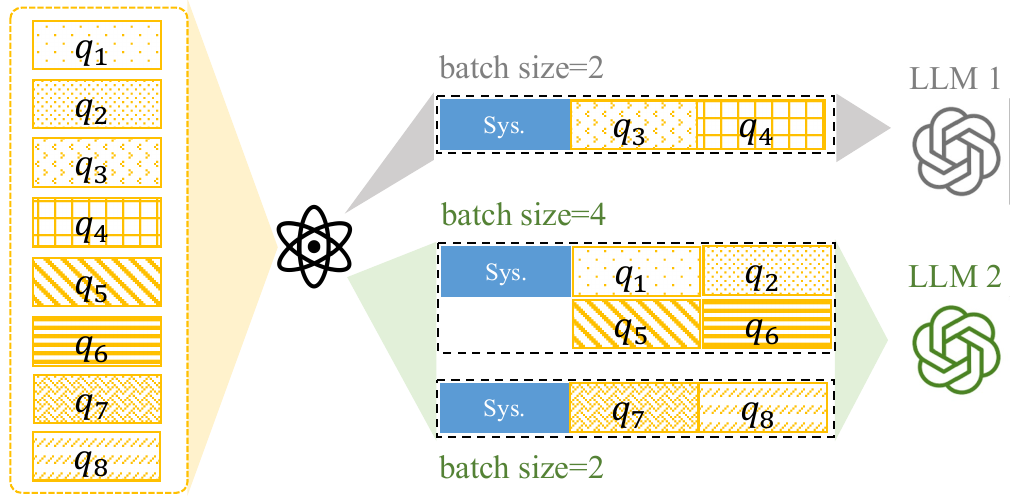}
			\label{fig:intro:route_and_batch}
		}
	\end{tabular}
	\caption{LLM Routing vs. Batch Prompting vs. Route with Batching}
    \label{fig:intro}
\end{figure*}

A typical LLM serving pipeline follows a simple paradigm that for each incoming query, the system constructs a prompt comprising the task-specific system prompt followed by the query, and submits the full prompt to a target model. 
In practice, instead of relying on a single model, LLM serving systems often maintain a pool of models with different capabilities and prices~\cite{DBLP:journals/corr/abs-2601-10088, DBLP:journals/pacmmod/ZeighamiSP25, DBLP:journals/corr/abs-2601-05536, DBLP:journals/pvldb/PatelJPGAGZ25}. Smaller models are cheaper and faster, and often sufficient for simple queries; larger models are more expensive but more reliable for difficult reasoning or knowledge-intensive queries. 
This heterogeneity creates an opportunity for LLM routing~\cite{DBLP:conf/emnlp/HuangLLCZWL25}: if the system can identify, for each query, the cheapest model that achieves quality comparable to the best available model, it can reduce cost while preserving overall quality of the answers. 
The key idea of LLM routing is to learn a routing strategy that predicts the most appropriate model for each query, as shown in Fig.~\ref{fig:intro:route}. 
For example, \FrugalGPT~\cite{DBLP:journals/tmlr/ChenZ024} and \RouteLLM~\cite{DBLP:conf/iclr/OngAWC0GKS25} send an easy query to cheaper models and reserve stronger and more expensive models for hard queries via query-level model selection. 
However, routing only optimizes the dimension of model assignment for LLM serving. It still invokes the LLM once per query, and thus leaves the repeated fixed cost of the system prompt paid again and again across the query workload. 

A second line of work called batch prompting~\cite{DBLP:conf/emnlp/ChengK023} addresses this cost-inefficiency. Rather than issuing an LLM call per query, as illustrated in Fig.~\ref{fig:intro:batch}, batch prompting packs multiple independent queries into an LLM invocation so that they share one common system prompt. 
Therefore, this amortizes the fixed input cost of the system prompt and can substantially reduce per-query cost. 
Recent frameworks, such as \BATCHER~\cite{DBLP:conf/icde/FanHFC00024} and \OBP~\cite{DBLP:journals/pvldb/JiWLXZ25}, explore grouping queries into batches by semantic similarity, diversity, or adaptive clustering. Intuitively, the longer the system prompt and the larger the batch size, the more cost expenditure can be saved.
However, batching also introduces its own trade-off between batch size and result quality. 
As the batch sizes grow, the result quality can degrade, especially when the input prompt exceeds the effective context window of the model~\cite{DBLP:conf/iclr/LinDDA24}. 

These two techniques, LLM routing and batch prompting, are logically complementary. 
Routing optimizes which model processes each query, operating along the dimension of model assignment, while batching optimizes how multiple queries share an LLM invocation, operating along the dimension of query aggregation. The real deployment lies not in either technique alone, but in their joint optimization.
Unfortunately, existing approaches only explore one side of this decision space. Routing methods assume non-batched prompting and estimate query performance under single-query invocations, as shown in Fig.~\ref{fig:intro:route}, whereas batching methods typically assume a fixed model and optimize only how queries are grouped, as shown in  Fig.~\ref{fig:intro:batch}.
A natural idea is to combine them in one pipeline:  first perform routing, then batch the queries assigned to each model; or first construct batches, then choose a model for each batch.
Another straightforward extension is to equip the pipeline with dynamic, adaptive batch sizes for each model. 
However, these adaptations are only coarse-grained and suboptimal, treating one technique as a loosely coupled add-on to the other instead of jointly optimizing the two-dimensional decisions.

The core difficulty of the joint optimization is that, once batching is introduced, the cost and utility of a configuration can no longer be characterized by the model alone.
Instead, they are jointly determined by the model, the batch size, and even the workload in a complicated way. 
Specifically, a larger and more expensive model prompted with a larger batch size may spend either less or more monetary cost per query than a smaller and cheaper model prompted with a smaller batch size. 
Likewise, a larger model with a larger batch size may achieve either lower or higher utility than the smaller model with a smaller batch size. As a result, this intricate and intertwined effect of model choice and batch size makes naive composition fundamentally inadequate, where either the router is batch-unaware, or the batching strategy is model-unaware.

This motivates a new perspective in which routing and batching should be treated as a single decision problem, where each query is assigned to an execution state specifying both the model choice and the batch size.
From this perspective, the serving system must re-profile a joint cost-utility landscape spanning both the model space and the batch size space and then determine an effective state assignment over the landscape. 
This task is challenging for three reasons.
First, the decision space is inherently combinatorial, and exhaustively profiling utility for every model–batch-size combination, especially in a workload-specific manner, is prohibitively expensive.
Second, the system should determine a principled candidate batch-size set for each model. 
Third, the state assignment algorithm must remain efficient for practical online deployment.

To address these challenges, we propose a framework for cost-effective LLM serving that jointly optimizes routing and batch prompting.
We first formulate the \emph{Route with Batching Problem} under a budget limit as a constrained combinatorial optimization problem and prove that it is NP-hard. 
With an in-depth analysis of the impact of model routing and batch prompting on cost and utility, we propose an integrated solution dubbed \RoBatch, which resolves the problem in two stages: modeling and routing. In the modeling stage, \RoBatch constructs a batch-aware proxy utility model that captures how each model’s utility varies with batch size. 
To efficiently learn a proxy model over the combinatorial space of models and batch sizes, \RoBatch decomposes the modeling into two sub-tasks: building a router that estimates query utility under non-batched execution, and fitting model-specific scaling functions that characterize the relative utility degradation as batch size increases. 
In the routing stage, \RoBatch treats each model--batch-size pair as a candidate state for a query, and performs fine-grained query-level scheduling over the candidate space, as Fig.~\ref{fig:intro:route_and_batch} displays. 
Specifically, we propose a greedy algorithm that progressively climbs the Pareto frontier of all queries under the budget constraint by upgrading the most cost-effective states. 
We conduct comprehensive experiments on six benchmarks across two LLM families, Qwen3 and Gemma3, comparing \RoBatch with four adapted routing and batching baselines.
The results demonstrate that \RoBatch consistently achieves a better Pareto frontier in the cost-utility landscape, while incurring modest scheduling overhead.


\stitle{Roadmap.} 
The rest of the paper is organized as follows. \cref{sec:bg} presents the background on LLM routing and batch prompting together with empirical observations that motivate our study. 
\cref{sec:formulation} formulates the route with batching problem and overviews our solution \RoBatch. \cref{sec:modeling} and \cref{sec:routing} elaborate on the modeling stage and routing stage of \RoBatch, respectively. \cref{sec:exp} reports the experimental studies. We review the related works in \cref{sec:rw} and conclude the paper in \cref{sec:conclusion}.

\section{Background \& Empirical Studies}
\label{sec:bg}

In this section, we first provide background on LLM routing and batch prompting, and then present respective empirical observations that motivate our study.

\subsection{LLM Routing \& the Impact}
\label{sec:bg:routing}

LLM routing is a model-serving technique that improves the cost-performance trade-off by dynamically assigning each query to the cheapest model predicted to answer it correctly. In a conventional LLM serving pipeline, every incoming query is dispatched to a single, pre-selected model. This one-size-fits-all strategy is inherently suboptimal: a smaller, cheaper model may suffice for simple queries, whereas complex queries may demand a larger, more powerful, yet more expensive model. LLM routing addresses this inefficiency with a learned router that examines, for each incoming query, dynamically assigns it to the most suitable model from a model pool $\mathcal{M} =\{m_1, \cdots, m_K\}$. 
The router is typically trained on historical query–result pairs to estimate the probability that each candidate model will correctly resolve a given query. At inference time, the router's prediction determines the target model for each query. By directing each query to the cheapest model expected to answer it correctly, routing can substantially reduce total monetary cost while preserving, or even improving, the overall task performance.

\begin{figure}[t]
     \begin{tabular}[h]{c}
        \subfigure[\AGNews] {
				\includegraphics[width=0.45\columnwidth]{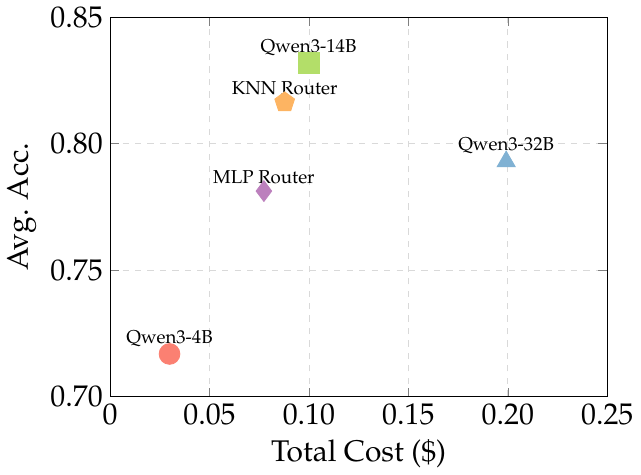}
			\label{fig:obs:router:agnews}
		} 
        \subfigure[\GSM] {
			\includegraphics[width=0.42\columnwidth]{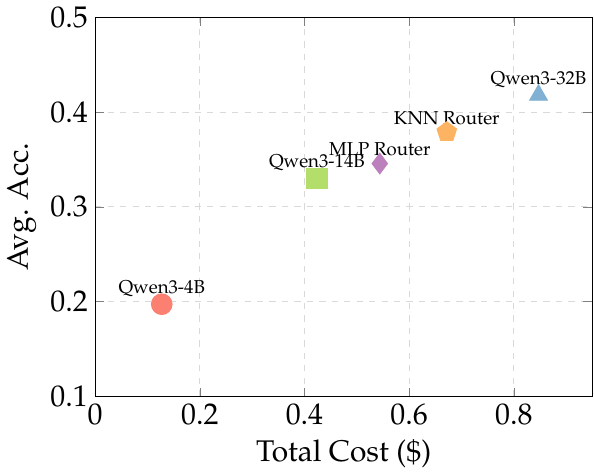}
			\label{fig:obs:router:gsm}
		}
	\end{tabular}
	\caption{The Impact of Routing on Avg. Acc. and Cost}
    \label{fig:obs:router}
\end{figure}

\stitle{Impact on Query Accuracy and Monetary Cost.} 
To explore the impact of routing on accuracy and cost, we compare two vanilla routers, a Multi-layer Perceptron (MLP)-based router, and a K-nearest neighbor (KNN)-based router, with the single-model baselines in the Qwen3 family. Fig.~\ref{fig:obs:router} presents each strategy as a point in the cost-accuracy space, on a topic classification query set \AGNews and a mathematical reasoning query set \GSM. 
In general, routing consistently achieves favorable trade-offs in the cost-accuracy landscape.
Simultaneously, by saving the costs from easier queries, routers selectively spend the saved budget on expensive models for hard queries, yielding a controllable mechanism under a fixed total budget. 
It is worth mentioning that the larger the model, the higher the cost, but larger models do not universally dominate the Accuracy for all tasks, which depends on task-specific characteristics. 



\begin{figure}[t]
     \begin{tabular}[h]{c}
        \subfigure[\AGNews] {
		      \includegraphics[width=0.45\columnwidth]{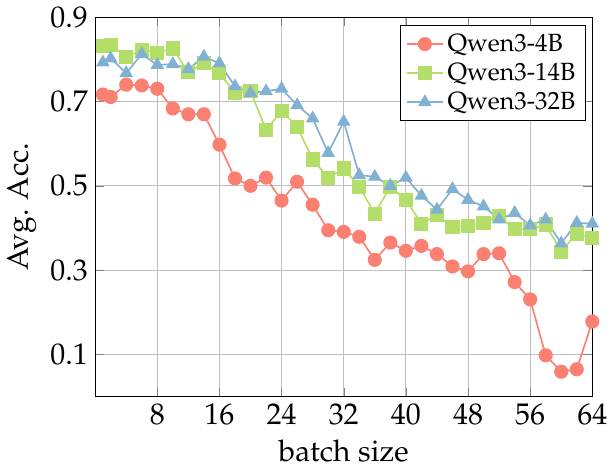}
			\label{fig:obs:batch:ass:agnews}
		} 
        \subfigure[\GSM] {
			\includegraphics[width=0.45\columnwidth]{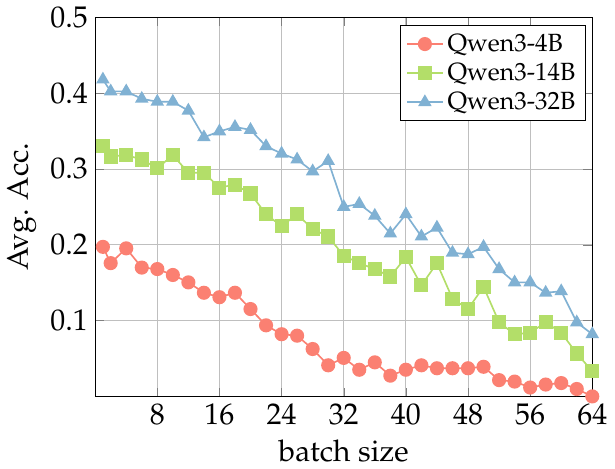}
			\label{fig:obs:batch:acc:gsm}
		}
	\end{tabular}
	\caption{The Impact of Batching on Avg. Acc.}
    \label{fig:obs:batch:acc}
\end{figure}

\subsection{Prompt Batching \& the Impact}
\label{sec:bg:batching}
Batch prompting reduces the cost of LLM invocations by packing multiple independent queries into a single context of invocation. 
In standard non-batched invocations, every invocation pairs the system prompt, comprising task instructions, formatting guidelines, and optional few-shot demonstrations with exactly one user query. 
Because the system prompt is identical across queries, its token cost is incurred repeatedly, constituting a fixed and structurally redundant overhead that scales linearly with the number of queries. 
Batch prompting amortizes this redundant cost by concatenating $b$ queries with a single, shared system prompt in one invocation.

\stitle{Impact on Query Accuracy and Monetary Cost.}
To investigate the impact of batch size $b$ on result quality and cost for different LLMs, we gradually increase $b$ from 1 to 64 by randomly grouping queries and report the average Accuracy of the three Qwen3 LLMs on the two datasets in Fig.~\ref{fig:obs:batch:acc}.
We observe that when the batch size is small, e.g., $b < 16$ for \AGNews and $b < 8$ for \GSM, accuracy remains stable or declines only marginally compared with single-query invocations (i.e., $b = 1$), with the tolerance depending on the tasks and the capacity of LLMs.
This stability persists as long as the concatenated input remains within the LLM's reasoning capability and its effective attention span, a.k.a., effective context length. 
However, as the batch size continues increasing, the excessive complexity of the concatenated prompt overwhelms the model's reasoning capability and effective context length, and the Accuracy drastically degrades to a low regime. 
This degradation is especially severe for the smaller Qwen3-4B, while the larger models, such as Qwen3-14B and Qwen3-32B, exhibit greater resilience.

\begin{figure}[t]
     \begin{tabular}[h]{c}
         \subfigure {
		      \includegraphics[width=0.95\columnwidth]{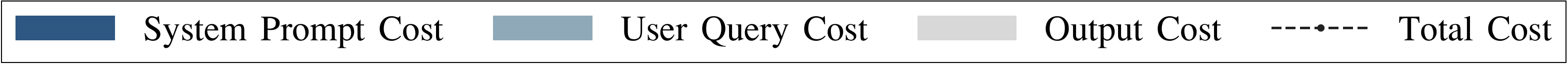}

		} \\
        \subfigure[\AGNews] {
		      \includegraphics[width=0.95\columnwidth]{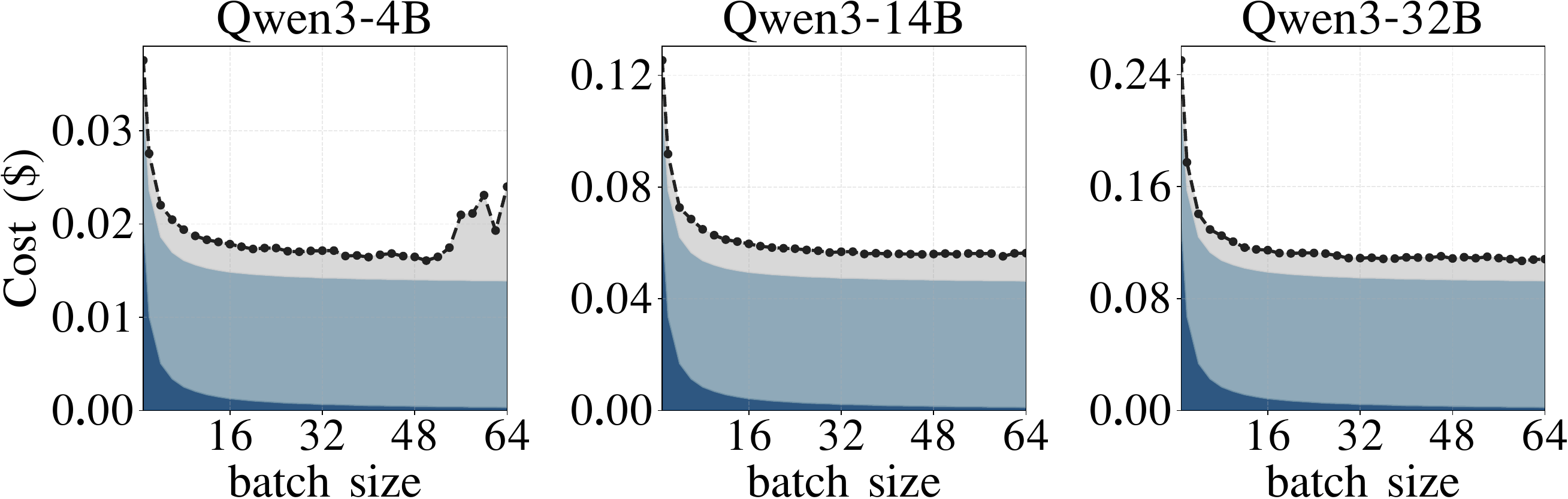}
			\label{fig:obs:batch:cost:agnews}
		} \\
        \subfigure[\GSM] {
			\includegraphics[width=0.95\columnwidth]{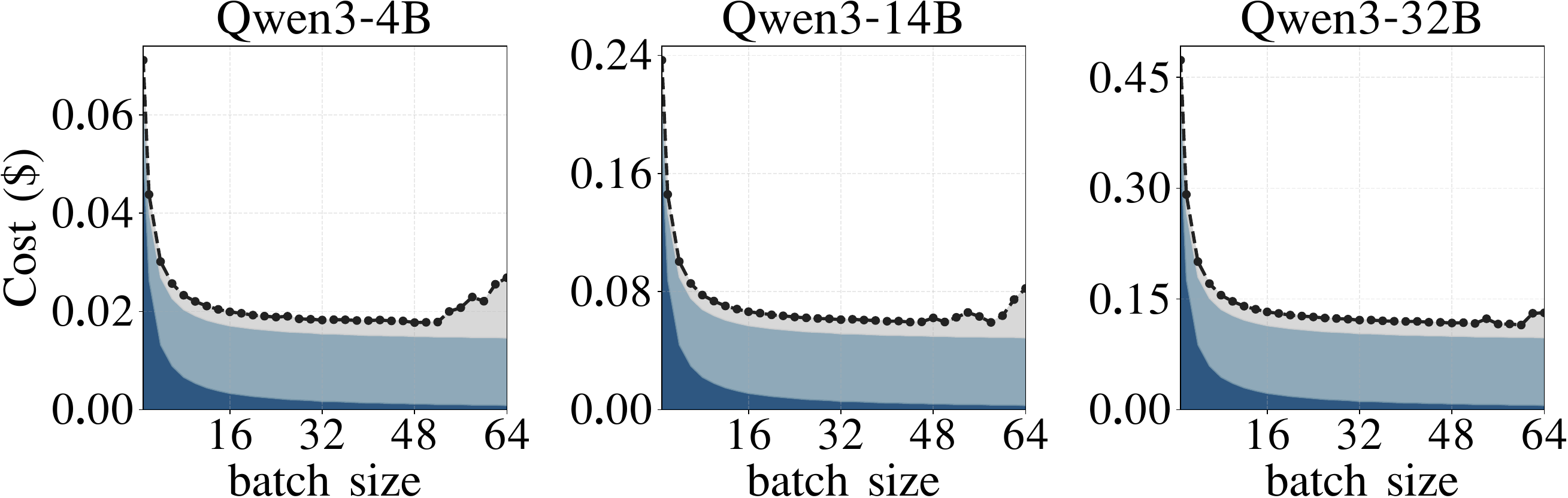}
			\label{fig:obs:batch:cost:gsm}
		}
	\end{tabular}
	\caption{The Impact of Batching on Cost}
    \label{fig:obs:batch:cost}
\end{figure}

We further investigate the impact of different batch sizes on the cost. Fig.~\ref{fig:obs:batch:cost} shows the total cost of system prompt, query, and the generated output across the three Qwen3 LLMs. 
We observe that the cost of query prompt and output remains stable except in the cases of $b>50$ on Qwen3-4B and larger batch sizes on the reasoning-intensive task \GSM where inference degeneration causes the LLM to produce repetitive or malformed outputs.
As $b$ increases from 1, the cost of system prompt is amortized by a factor of $1/b$. 
When $b=1$, the cost of system prompt occupies around 59.5\% and 90.1\% total cost for \AGNews and \GSM, respectively, while for $b= 16$ on \AGNews and $b=8$ on \GSM, the shares are amortized to $8.4\%$ and $53.2\%$, respectively, demonstrating that the amortization results in a significant cost reduction. 

\subsection{Takeaway \& Motivations}
The empirical studies in \cref{sec:bg:routing} and \cref{sec:bg:batching} shed light on two complementary insights. First, routing optimizes \emph{which} model processes each query, but not \emph{how} queries are processed. 
LLM routing provides a controllable mechanism for navigating the cost–accuracy Pareto frontier at the granularity of individual queries. However, routing processes each query as an individual LLM invocation, leaving the structurally redundant cost from the repetitive system prompt unoptimized, which grows linearly with the number of queries, regardless of the routing strategy.
Second, batch prompting optimizes \emph{how} queries share an invocation, but not \emph{which} model they are assigned to. 
Batch prompting amortizes the cost of the system prompt by packing multiple queries into one LLM invocation. 
Exceeding a model's effective batch size degrades accuracy and may inflate costs, indicating that the batch size should be carefully calibrated for each model based on the specific task. 

\stitle{Motivation.} 
These two findings reveal a fundamental complementarity: routing controls the model choice dimension $m_k$ while batching controls the query aggregation dimension $b$, but neither technique alone explores the joint optimization space $(m_k, b)$. 
More significantly, the two dimensions interact non-trivially.
As shown in Fig.~\ref{fig:obs:batch:acc}, the effective batch size depends on both model capability and task characteristics. 
The cost reduction from batching changes the per-query pricing of each model, and the accuracy degradation from batching affects different models by different degrees, which jointly reshape the cost-accuracy landscape. 
This motivates us to a unified formulation where model selection and batch size assignment are jointly optimized.

\section{Problem Formulation \& Overview}
\label{sec:formulation}
We use $\mathcal{Q} = \{q_1, \cdots, q_n\}$ to represent $n$ queries to be solved by LLMs, and use $\mathcal{M} = \{m_1, \cdots, m_K\}$ to represent a pool of $K$ LLMs. Each LLM $m_k \in \mathcal{M}$ is associated with distinct per-token costs for input and output, denoted as $c^{in}_k$ and $c_k^{out}$, respectively.
Considering solving a single query $q_i$ by LLM $m_k$ in one LLM invocation, we characterize the monetary cost of this invocation as Eq.~\eqref{eq:singlecost}:
\begin{align}
    C_{\text{sys}}(m_k) &+ C_{q_i}(m_k), \label{eq:singlecost} \\
    C_{\text{sys}}(m_k) &= t_\text{sys}\cdot c^{in}_k, ~~C_{q_i}(m_k) = t^{in}_{q_i} \cdot c^{in}_k + t^{out}_{q_i} \cdot c^{out}_k
\end{align}
where $C_{\text{sys}}(m_k)$ is the system prompt cost, determined by a fixed number of tokens $t_\text{sys}$, and $C_{q_i}(m_k)$ is the cost incurred purely by the query $q_i$, including the cost of $t^{in}_{q_i}$ tokens from the input query and  $t^{out}_{q_i}$ tokens from the generated output. In this paper, we assume that the number of output tokens is easy to estimate, e.g., the response of a multi-class classification problem where output tokens are fixed, or by restricting the largest number of output tokens in the system prompt. 
Without loss of generality, we assume the $K$ models in $\mathcal{M}$ are indexed in ascending order of size and capability, i.e., $m_1 < \cdots < m_K$, with their per-token costs ordered $c^{in}_1 < \cdots < c^{in}_K$ and $c^{out}_1 < \cdots < c^{out}_K$ accordingly. 

Batch prompting packs multiple queries into a single LLM invocation. 
Given the query workload $\mathcal{Q}$, each LLM $m_k$ is characterized by a set of valid batch sizes $\mathcal{B}_k = \{1, \cdots, b^\text{max}_k\}$ in ascending order where $b^\text{max}_k$ is the largest batch size, determined by the average query length in $\mathcal{Q}$ and the effective context length of $m_k$. 
When a batch size $b \in \mathcal{B}_k$ is adopted, the system prompt of $C_\text{sys}(m_k)$ is amortized by $1/b$ for each query, while the per-query cost $C_{q_i}(m_k)$ remains unchanged. 
Consider that a query in $\mathcal{Q}$ is simultaneously routed to an LLM in $\mathcal{M}$ and assigned a batch size from the space of the batch size of that LLM.
We assign each query a state $s(q_i) = (m_k, b)$, denoting that $q_i$ will be routed to LLM $m_k$ and  packed into a batch with a size of $b \in \mathcal{B}_k$. A binary decision variable $x_{i, k, b} \in \{0, 1\}$ for $i \in \{1, \cdots, n\}, k \in \{1, \cdots, K\}, b \in \mathcal{B}_k$ encodes this assignment:
\begin{align}
    x_{i, k, b} = 1 \Longleftrightarrow s(q_i) = (m_k, b), \text{and}~0~\text{otherwise}.
\end{align}
Given a specific assignment of the decision variables $x_{i, k, b}$, the total cost from model $m_k$ under batch size $b$ is 
\begin{align}
    C(m_k, b) = \left\lceil \frac{1}{b} \sum_{i=1}^{n} x_{i,k,b} \right\rceil \cdot C_\text{sys}(m_k) + \sum_{i = 1}^n x_{i,k,b} \cdot C_{q_i}(m_k). \label{eq: model_batch_cost}
\end{align}
The first term of Eq. (4) captures the total system prompt cost across all batches of size $b$, where the ceiling operator $\lceil \cdot \rceil$ accounts for the number of physical invocations, 
including partially filled batches. The second term captures the cumulative input and output token cost of all queries assigned to the state $(m_k, b)$.


It is worth noting that a single LLM may simultaneously serve batches of different sizes. As shown in Fig.~\ref{fig:obs:router} and Fig.~\ref{fig:obs:batch:acc}, query accuracy is influenced by both the routed model and the batch size. We therefore let $u_{i,k,b}$ denote the utility score of query $q_i$ in state $(m_k, b)$, e.g., whether $q_i$ is correctly classified in a classification task.
Given a monetary cost budget $C_{bdg}$, the cost-effective LLM routing with batch prompting is formally defined as the following constrained combinatorial optimization problem: 
\begin{definition}
(\textbf{Route with Batching Problem})
Given a query workload $\mathcal{Q} = \{q_1, \cdots, q_n\}$, a set of LLMs $\mathcal{M} = \{m_1, \cdots, m_K\}$, and a total cost budget $C_{bdg}$. Suppose each model $m_k$ is associated with a set of valid batch sizes $\mathcal{B}_k$. The problem is to seek a binary assignment $x_{i, k, b} \in 
\{0, 1\}$ that solves:
\begin{align}
   \max \sum_{i=1}^{n} \sum_{k=1}^{K} \sum_{b \in \mathcal{B}_k} x_{i,k,b} \cdot u_{i,k,b}, \label{eq:prob:utility} \\
   s.t.  \sum_{k=1}^{K} \sum_{b \in \mathcal{B}_k} x_{i,k,b} = 1~~\forall i \in \{1, \dots, n\}, \label{eq:prob:constraint1} \\
   \sum_{m_k \in \mathcal{M}} \sum_{b \in \mathcal{B}_k} C(m_k,b)\le C_{bdg}. \label{eq:prob:constraint2}
\end{align}
The objective (Eq.~\eqref{eq:prob:utility}) is to maximize the total utility of all processed queries, subject to two constraints: Eq.~\eqref{eq:prob:constraint1} ensures that each query $q_i$ is routed to exactly one LLM with exactly one batch size, and Eq.~\eqref{eq:prob:constraint2} ensures that the total cost does not exceed the given budget.
\end{definition} 

\stitle{The NP-hardness of Route with Batching Problem.}
We analyze the computational hardness of the  problem as follows.
\begin{theorem}
The Route with Batching Problem is NP-hard.
\end{theorem}
\proofsketch
We reduce the maximum coverage (MC) problem~\cite{Karp72}, which is known to be NP-hard, to the route with batching problem.
Given an integer $B$ and a collection of sets where each set contains some elements, the MC problem is to find at most $B$ sets that cover the largest number of elements.
Consider an arbitrary instance $H$ of MC with $K$ sets $T_1,\cdots,T_K$ and $n$ elements $\{e_1,\cdots,e_n\}=\bigcup_{1 \le k \le K} T_k$. We construct a corresponding instance of the route with batching problem as follows.
Each query $q_i$ corresponds to an element $e_i$ $(1 \le i \le n)$, and each LLM $m_k$ corresponds to a set $T_k$ $(1 \le k \le K)$.
For each LLM $m_k$, we let its valid batch-size space contain only one batch size: $\mathcal{B}_k=\{n\}$. We set the system prompt cost of each model to $C_{\text{sys}}(m_k)=1$, and the query-specific cost to $C_{q_i}(m_k)=0$ for every $q_i$ and $m_k$. For each query-model pair, we define the utility as $u_{i,k,n}=1$ if $e_i \in T_k$ and $u_{i,k,n}=0$ otherwise. Additionally, we set the total budget to $C_{bdg}=B$.
For any model $m_k$, let $N_k$ be the number of queries assigned to $m_k$. Since the only valid batch size is $n$, the total cost incurred by $m_k$ is $C(m_k,n)=\left\lceil \frac{N_k}{n} \right\rceil \cdot C_{\text{sys}}(m_k)=\left\lceil \frac{N_k}{n} \right\rceil.$
Because $0 \le N_k \le n$, we have $C(m_k,n)=1$ if $N_k>0$, and $C(m_k,n)=0$ otherwise. Thus, the total routing cost is exactly the number of used models.
Consider any feasible solution to the constructed instance. Let $I$ be the set of used models. Since the total cost equals the number of used models and is bounded by $C_{bdg}=B$, we have $|I| \le B$. For each query $q_i$, its utility contribution is at most $1$, and it can achieve utility $1$ if and only if it is assigned to a model $m_k \in I$ such that $e_i \in T_k$. Therefore, the maximum total utility obtainable by the set $I$ is exactly the number of elements covered by the corresponding set collection $\{T_k \mid m_k \in I\}$. Hence, an optimal solution to the constructed routing instance yields an optimal solution to the MC instance, and vice versa. Since the MC problem is NP-hard, the route with batching problem is NP-hard.
\eop

\stitle{The Overview of \RoBatch. }
Our cost-effective routing with batch prompting framework, \RoBatch, operates in two stages: a modeling stage that constructs a batch-aware proxy model for each LLM in $\mathcal{M}$ and a routing stage that employs a greedy algorithm to assign each query $q_i \in \mathcal{Q}$ to a state $s(q_i) = (m_k, b)$ based on the batch-aware proxy model. 
In the modeling stage, the principal task of \RoBatch is to estimate query utility in each LLM $m_k$ regarding different batch sizes in the valid space of $\mathcal{B}_k$, profiling the landscape of query utility changes regarding different models and batch sizes. 
Recall that typical LLM routing approaches estimate query utility under a batch size of 1, whereas \RoBatch needs to further account for the impact of batching. 
However, directly constructing a proxy model that jointly models the query space, the model space, and the batch size space faces a combination explosion. 
Thereby, \RoBatch introduces the batch-aware proxy model as Eq.~\eqref{eq:proxy_utility}, decoupling the proxy model for single query invocation with a query-agnostic scaling function parametrized by batch size:

\begin{align}
    \widehat{u}_{i, k, b} = \widehat{u}_{i, k, 1} \cdot \rho_k(b), \label{eq:proxy_utility}
\end{align} 
where $\widehat{u}_{i, k, b}$ is the estimated utility score of query $q_i$ from LLM $m_k$ with  batch size $b \in \mathcal{B}_k$, and $\widehat{u}_{i, k, 1}$ is the estimated utility of $q_i$ from LLM $m_k$ without batching, i.e., $b=1$.
The term $\rho_k(b)$ is a model-specific scaling function that captures the general relative decay in  utility as the batch size increases from $1$ to $b$.  
During the construction of the scaling function $\rho_k(b)$ and the proxy model $\widehat{u}_{i, k, b}$, \RoBatch also calibrates the effective batch size $b^\text{effect}_k$ for each LLM $m_k$ as the maximum batch size in $\mathcal{B}_k$ to use.
We provide the details of the modeling stage in \cref{sec:modeling}. 
In the routing stage, \RoBatch begins by assigning every query an initial state, which is the cheapest LLM in the model pool with that model's effective batch size. Then, \RoBatch iteratively upgrades the states of queries in a greedy manner, prioritizing queries whose per-unit-cost utility can be improved by the largest margin. The iteration continues until either the total cost budget is exhausted or all queries have reached their best states. We elaborate on the greedy algorithm in \cref{sec:routing}.

\section{The Modeling Stage}
\label{sec:modeling}

The modeling stage constructs a batch-aware proxy model for each LLM $m_k$, which estimates the utility score of query $q_i$ from $m_k$ with a batch size of $b$, denoted as $\widehat{u}_{i, k, b}$. Constructing such a model poses two challenges. First, the LLM inference is context-dependent; the practical utility of a query $q_i$ is also influenced by the other $(b-1)$ queries sharing the same invocation, which cannot be determined beforehand. Second, jointly modeling the query space, the model space, and the batch size space leads to a combinatorial explosion. 
To address these challenges, \RoBatch adopts a simple yet effective formulation based on the observation that queries in one workload are different problem instances of one general task. 
Consequently, as $b$ increases from 1, different queries tend to exhibit a similar trend of utility degradation on an LLM $m_k$. 
\RoBatch factorizes this trend into a query-agnostic scaling function $\rho_k(b)$, separating it from the utility score without batching $\widehat{u}_{i, k, 1}$, as shown in Eq.~\eqref{eq:proxy_utility}.  The scaling function $\rho_k(b)$ is expected to also encapsulate the complex influence of the $b-1$ co-batched random queries. As a result, estimating $\widehat{u}_{i, k, b}$ reduces to independently estimating $\rho_k(b)$ and $\widehat{u}_{i, k, 1}$. We introduce how each is obtained below.

\stitle{Fitting the Utility Scaling Function. } Profiling the scaling function $\rho_{k}(b)$ for each LLM $m_k$ is a novel and critical task in  \RoBatch. Before fitting $\rho_{k}(b)$, \RoBatch first calibrates two quantities for each LLM: the largest batch size $b^\text{max}_k$ and the effective batch size $b^\text{effect}_k$. Here, $b^\text{max}_k$ serves as the upper limit for the calibration of $b^\text{effect}_k$ and $b^\text{effect}_k$ is the most important hyper-parameter for batch prompting, as it further determines the upper bound of the valid batch size space $\mathcal{B}_k$ of model $m_k$. 
Both quantities depend on the query workload $\mathcal{Q}$, but in different ways. $b^\text{max}_k$ is determined by the average query token length and the system prompt length, while $b^\text{effect}_k$ is jointly governed by the query difficulty and the capability of model $m_k$.
To derive the largest batch size $b^\text{max}_k$, \RoBatch introduces a threshold $\epsilon \in (0, 1)$ that bounds the fraction of system prompt cost in the batched prompt for each model $m_k$:
\begin{align}
\epsilon \leq \frac{C_\text{sys}(m_k)}{{C_\text{sys}(m_k)} + b_k \cdot \mathbb{E}_{q_i \in \mathcal{Q}} [{C_{q_i}}(m_k)]}, \label{eq:batchsize:bound}
\end{align}
where $C_\text{sys}(m_k)$ is the system prompt cost, ${C_{q_i}}(m_k)$ is the cost of an individual input query, and $\mathbb{E}_{q_i \in \mathcal{Q}} [{C_{q_i}}(m_k)]$ is the expected per-query cost over  $\mathcal{Q}$. Rearranging the inequality of Eq.~\eqref{eq:batchsize:bound} yields the upper bound on $b_k$:
\begin{align}
    b_k \leq b^{\max}_{k} = \left\lceil \frac{C_\text{sys}(m_k)(1 - \epsilon)}{\epsilon \cdot \mathbb{E}_{q_i \in \mathcal{Q}} [{C_{q_i}}(m_k)] } \right\rceil. \label{eq:batchsize:largest}
\end{align}
Given a threshold $\epsilon$, \RoBatch can compute $b^{\max}_{k}$ for each model $m_k$ directly via Eq.~\eqref{eq:batchsize:largest} without extra profiling. 
Subsequently, \RoBatch calibrates the effective batch size $b^\text{effect}_k$ in the range of $[1, b^{\max}_{k}]$ for each model $m_k$, respectively, and simultaneously fits the scaling function $\rho_k(b)$ over its valid batch size space $\mathcal{B}_k = [1, \cdots, b^\text{effect}_k]$. 
Given the significance of $b^\text{effect}_k$ and $\rho_k(b)$, we re-utilize the training query set $\mathcal{Q}'$ to profile each model's performance under batch prompting with a batch size $b \in \mathcal{B}_k$.

\begin{figure}[t]
  \centering
	\includegraphics[width=0.85\columnwidth]{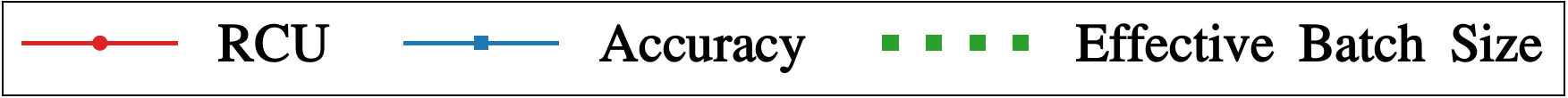}
\begin{tabular}[h]{c}
	\subfigure[\AGNews] {
  \includegraphics[width=1\linewidth]{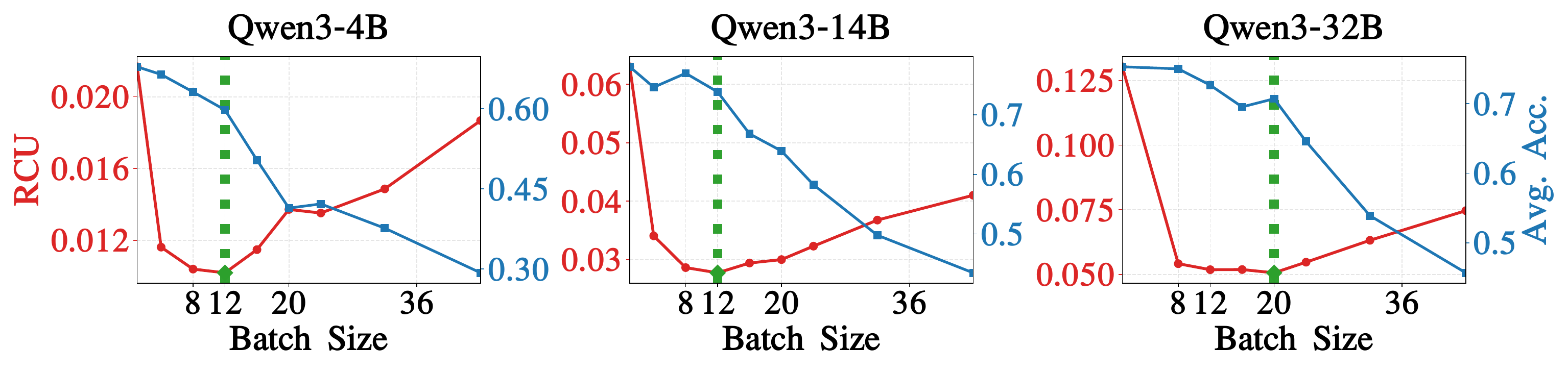}
  } \\
  	\subfigure[\GSM] {
  \includegraphics[width=1\linewidth]{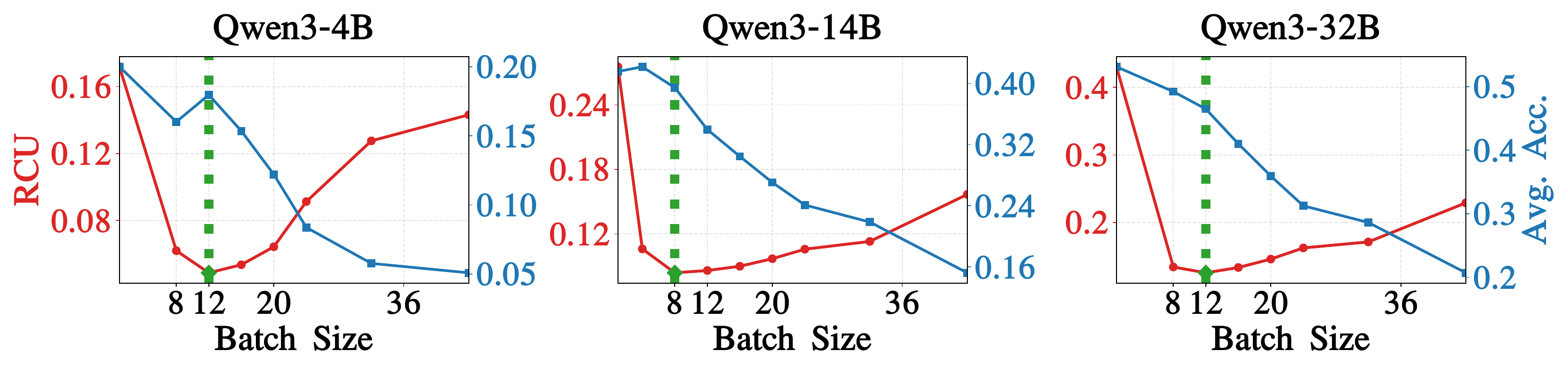}
  }
\end{tabular} 
  \caption{RCU and Avg. Acc. Curves of Different Batch Sizes. The Effective Batch Size Is Identified Via Ternary Search.}
  \label{fig::framework::cer_trend}
\end{figure}

Although the ground-truth utility of $\mathcal{Q}'$ under $b=1$ has already been obtained, evaluating the full-scale training set for every model $m_k$ in its remaining batch size space $\mathcal{B}_k \setminus \{1\}$ via LLM invocations is time and cost consuming.
To reduce this overhead, we first extract a coreset from $\mathcal{Q}'$ that serves as a compact summary of $\mathcal{Q}'$. 
Specifically, by applying the $k$-center algorithm~\cite{DBLP:journals/tcs/Gonzalez85} on the vectorized embeddings of $\mathcal{Q}'$, a smaller yet semantically diverse set of queries, $\mathcal{Q}''$, is extracted as the representative of the entire training set $\mathcal{Q}'$.
On the coreset $\mathcal{Q}''$, we search for a ‘sweet point' in the range $[1, b^{\max}_k]$ for each model $m_k$ as the effective batch size that minimizes the ratio of cost to utility (RCU) as below:
\begin{align}
    \text{RCU}(b) = \frac{{C_\text{sys}(m_k)} + b \cdot \mathbb{E}_{q_i \in \mathcal{Q}''} [{C_{q_i}}(m_k)]}{\mathbb{E}_{q_i \in \mathcal{Q}''} [u_{i, k, b}]}. \label{eq:rcu}
\end{align}
Here, the numerator of RCU in Eq.~\eqref{eq:rcu} is the expected cost of a batched prompt of size $b$ for model $m_k$, computed by averaging the cost over the coreset $\mathcal{Q}''$, while the denominator is the expected utility of that batched prompt for model $m_k$, computed by averaging the evaluated utilities over $\mathcal{Q}''$ via LLM invocations.
Intuitively, RCU quantifies the monetary cost spent on unit utility at a specific batch size $b$. \RoBatch uses it as a profiling criterion for selecting $b^\text{effect}_k$, since it follows an inherent consistency with our optimization objective (Eq.~\eqref{eq:prob:utility}) and constraint (Eq.~\eqref{eq:prob:constraint2}).
In general, as $b$ increases from 1 to $b^{\max}_{k}$, the RCU first decreases and then increases, exhibiting a ‘V'-shape. This behavior is because, at small batch sizes, the cost amortization from batching is substantial while the utility degrades only slightly. As $b$ continues to grow into a large regime, the cost amortization benefit becomes marginal while the utility drops significantly. Fig.~\ref{fig::framework::cer_trend} illustrates this trend of RCU (in red lines) on the two datasets, \AGNews and \GSM, using the Qwen3 family LLMs.
Since the RCU curve is uni-modal, 
\RoBatch performs an efficient ternary search over $[1, b^{\max}_{k}]$ to locate the batch size with minimal RCU for each model, as shown in Fig.~\ref{fig::framework::cer_trend}. This batch size is calibrated as the effective batch size $b^\text{effect}_k$. 
Intuitively, $b^\text{effect}_k$ is the most economical batch size for the workload within the effective context length of the LLM, a point at which the utility (e.g., Avg. Accuracy in the blue lines in Fig.~\ref{fig::framework::cer_trend}) is largely preserved. 

With $b^\text{effect}_k$ determined, we model the utility scaling function $\rho_k(b)$ regarding batch size $b\in [1, \cdots, b^\text{effect}_k]$. There are multiple ways to fit the decaying curves shown in Fig.~\ref{fig:obs:batch:acc} and Fig.~\ref{fig::framework::cer_trend}. 
By default, \RoBatch adopts piecewise linear interpolation on the utilities obtained from the training coreset $\mathcal{Q}''$, as it is numerically stable and parameter-free. 
Specifically, for each batch size $b_j \in [1, b_k^\text{effect}]$ and model $m_k$, we compute the scaling factor at $b_j$ as follows:
\begin{align}
\rho_k(b_j) = \frac{1}{u_{k,1}} \left( u_{k,b_{j-1}} + (b_j - b_{j-1}) \cdot \frac{u_{k,b_{j+1}} - u_{k,b_{j-1}}}{b_{j+1} - b_{j-1}} \right), \label{eq:linear_interp}
\end{align}
where $b_j$ and $b_{j+1}$ are the two nearest batch sizes such that $b_{j-1} \le b_j \le b_{j+1}$, and $u_{k, b}$ is the average utility of all queries in $\mathcal{Q}''$ by model $m_k$ at batch size $b$.
We investigate alternative formulations for fitting $\rho_k(b)$ in \cref{sec:exp:sensitivity}. 

\stitle{Estimation of the Utility Without Batching.} 
Existing routers serve as performance estimators that map a query to its estimated utility scores of all candidate models in $\mathcal{M}$ without batch prompting. 
To train such a router, we first establish a training dataset $\mathcal{Q}' = \{q'_1, \cdots, q'_n\}$. 
For each training query $q'_i \in \mathcal{Q}'$, we perform an offline evaluation by invoking all $K$ models with $b=1$ and collecting the real utility score $u_{i,k,1} \in \{0, 1\}$, which indicates whether model $m_k$ correctly resolves query $q'_i$.
This process generates the ground-truth binary vector $\bm{u}_i = [u_{i,k,1}, \cdots, u_{i,K,1}] $ for each $q'_i$. 
The input features of the router are vectorized embeddings of queries in $\mathcal{Q}'$, extracted offline by a sentence embedding model~\cite{DBLP:journals/corr/abs-2212-03533}. 
We formulate routing as a multi-label classification task and train an MLP or KNN-based classifier. The MLP classifier is trained by minimizing the binary cross entropy (BCE) loss over the training dataset $\mathcal{Q}'$.
Once trained, the classifier estimates the likelihood that each model $m_k \in \mathcal{M}$ can correctly resolve a test query $q_i$, i.e., $\widehat{u}_{i, k, 1} \in [0, 1]$. 
To minimize the overhead of router training and training query annotation, \RoBatch utilizes a simple three-layer MLP or KNN classifier. Studies on the design space of router architectures are orthogonal to the \RoBatch framework.

With the utility estimation for $b=1$ and the scaling value for $b \in [1, b^\text{effect}_{k}]$, \RoBatch combines them via multiplication as in Eq.~\eqref{eq:proxy_utility} to obtain utility estimations under batch prompting. 
These batch-aware estimations serve as the basis for decision-making during query routing.
We conclude this section by analyzing the complexity of the modeling stage.

\stitle{Complexity Analysis.} 
The offline modeling stage incurs an overall time complexity of $\mathcal{O}(|\mathcal{Q}'| |\mathcal{Q}''| + C_\text{API} \sum_{k=1}^K \log(b^{\max}_k) + |\mathcal{Q}'| d)$, where $C_\text{API}$ denotes the complexity introduced by LLM calls on the coreset data and $d$ denotes the dimension of the query embeddings. The overall complexity comprises three parts: coreset selection, model profiling, and router training.
First, coreset selection applies the $k$-center algorithm to extract a representative subset $\mathcal{Q}''$ from the full training set $\mathcal{Q}'$ in $\mathcal{O}(|\mathcal{Q}'||\mathcal{Q}''|)$ time. Due to $|\mathcal{Q}''| \ll |\mathcal{Q}'|$, this step remains efficient even for large-scale datasets. Second, \RoBatch calibrates the effective batch size $b^{\text{effect}}_k$ for each of the $K$ models by performing a ternary search over the batch size space $[1, b^{\max}_k]$. 
This search entails $\mathcal{O}(\sum_{k=1}^K \log b^{\max}_k)$ configuration evaluations, each of which requires $\lceil |Q''|/b_k \rceil$ LLM invocations at a total complexity of $C_\text{API}$, contributing $\mathcal{O}(C_\text{API} \sum_{k=1}^K \log b^{\max}_k)$ to the total complexity. The third part, router training, trains the router on the full training dataset. This step scales linearly with the dataset size and embedding dimensionality, requiring $\mathcal{O}(|\mathcal{Q}'| d)$ time.

\section{The Routing Stage}
\label{sec:routing}

\begin{figure}[t]
  \centering
  \includegraphics[width=0.8\linewidth]{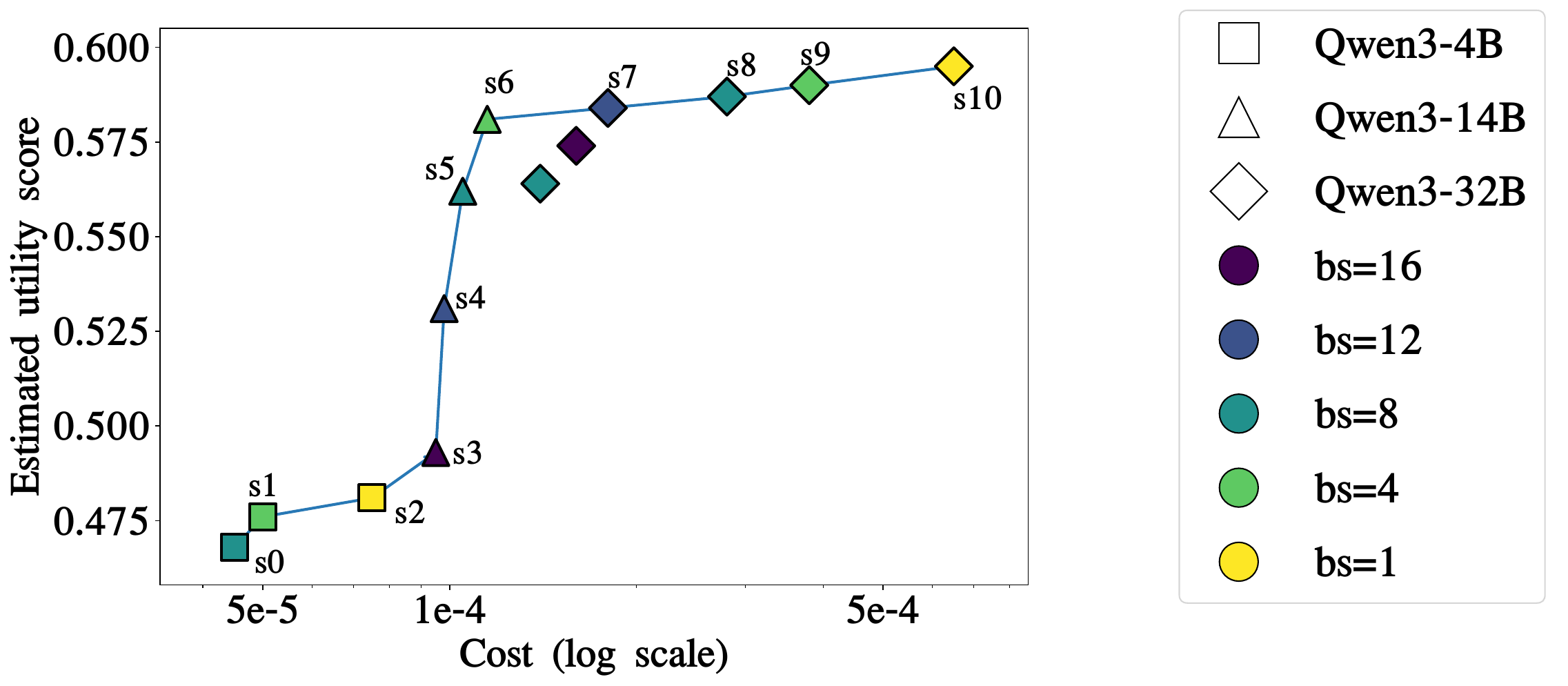}
  \caption{The Candidate States and Pareto Frontier of a Query}
  \label{fig:pareto_frontier}
\end{figure}

With the proxy utility $\widehat{u}_{i, k, b}$ for batch prompting over each model $m_k$ prepared, in the routing stage, \RoBatch assigns each test query in $\mathcal{Q}$ to a state $(m_k, b)$, thereby constructing a scheduling mapping $\mathcal{S}: q_i \mapsto (m_k, b)$, where $ b\in \mathcal{B}_k$. For each query $q_i$, unlike standard routing without batching that only considers $K$ candidate models, the batch-aware setting introduces $\sum_{k=1}^{K} |\mathcal{B}_k|$ candidate states in total. 
We first define a dominance relation of states and then show that it can be exploited to prune candidate states.
\begin{definition}
\label{def:dominace}
(\emph{Dominance of States}) Given a query $q_i$, let $s = (m_{k}, b)$ and $s' = (m_{k'}, b')$ denote two candidate states. Let $C_{q_i}(s)$ and $C_{q_i}(s')$ denote the cost of $q_i$ in states $s$ and $s'$, respectively, where: 
\begin{align}
    C_{q_i}(s) = C_\text{sys} / b + C_{q_i}(m_k), C_{q_i}(s') = C_\text{sys} / b' + C_{q_i}(m_{k'}).  \label{eq:per-query-cost}
\end{align}
Formally, state $s$ is dominated by $s'$ on $q_i$, denoted as $s \preceq_{q_i} s'$, \emph{iff} (1) $C_{q_i}(s) \geq C_{q_i}(s')$ and (2) $\widehat{u}_{i, k, b} \leq \widehat{u}_{i, k', b'}$.
\end{definition}
For a query $q_i$, a state $s$ is called a dominated state if there exists a state $s' \neq s$ in the candidate space satisfying $s \preceq_{q_i} s'$; otherwise, $s$ is an undominated state of $q_i$. 
All the undominated states of $q_i$ form the Pareto frontier in the cost-utility landscape, and dominated states can be pruned during scheduling. Fig.~\ref{fig:pareto_frontier} illustrates an example where each node represents a candidate state in the cost-utility space. The non-dominated nodes form the Pareto frontier shown as the blue line, and the remaining two nodes are dominated. 
We will discuss the safety of this pruning later in this section. 

\RoBatch employs a greedy algorithm to perform routing with batch prompting. Initially, every query $q_i \in \mathcal{Q}$ is assigned to the  initial state $s^{(0)} := (m_{1},  b^\text{effect}_1)$, where $m_{1}$ is the cheapest model in the model pool $\mathcal{M}$, and $b^\text{effect}_1$ is its effective batch size.
Notably, for any query, the initial state $s^{(0)}$ always lies on the cost-utility Pareto frontier since it attains the globally lowest cost $C_{q_i}(s^{(0)})$ among all candidate states. 
The scheduling algorithm then iteratively upgrades query states along the Pareto frontier by allocating more cost, until either the cost budget $C_{bdg}$ is exhausted or all queries reach their highest-utility states.
To formalize this process, let $[s^{(0)}, s^{(1)}, \cdots, s^{(T)}]$ denote the sequence of non-dominated states of query $q_i$, ordered by ascending per-query cost i.e., $C_{q_i}(s^{(0)}) < C_{q_i}(s^{(1)}) \leq \cdots C_{q_i}(s^{(T)})$. 
In each iteration, the algorithm selects the query $q_i$ whose next state upgrade yields the maximum utility gain per unit cost, $\Delta_{q_i}^{(t)}$, defined as:
\begin{align}
    \Delta_{q_i}^{(t+1)} = \frac{\widehat{u}_{i, s^{(t+1)}} - \widehat{u}_{i, s^{(t)}} }{ C_{q_i}(s^{(t+1)}) - C_{q_i}(s^{(t)}) }, \label{eq:rcu_slope}
\end{align}
where, with a slight abuse of notation, $\widehat{u}_{i, s^{(t)}}$ denotes the estimated utility of $q_i$ in state $s^{(t)} = (m_k, b)$, i.e., $\widehat{u}_{i, s^{(t)}} = \widehat{u}_{i, k, b}$, as predicted by the budget-aware proxy utility model constructed in \cref{sec:modeling}. 
In other words, the algorithm greedily allocates the remaining budget to the query whose utility can be improved to the greatest extent per unit of cost. 
An acute reader may find that $\Delta_{q_i}^{(t+1)}$ is the discrete slope of the inverse of the RCU (Eq.~\eqref{eq:rcu}) along the Pareto frontier at the query level. In essence, the algorithm optimizes the total utilities by following the direction of steepest RCU improvement, while ensuring that the cumulative cost remains within the budget constraint.

\begin{algorithm}[t]
\caption{Greedy Scheduling}
\small
\label{alg:greedy_upgrade}
\KwIn{queries $\mathcal{Q} = \{q_1, \dots, q_n\}$, total budget $C_{bdg}$, proxy utility model $\widehat{u}_{i, k, b}$}
\KwOut{Final State Assignment Mapping $\mathcal{S}: q_i \to (m_k, b)$}

Initialize Priority Queue $PQ \gets \{\}$ \;  
\For{$q_i \in \mathcal{Q}$}{ \label{line:init:start}
    $s^{(0)} \gets (m_1, b^\text{effect}_1)$ \; 
    $\mathcal{S}(q_i) \gets s^{(0)}$\;
    $C_{bdg} \gets C_{bdg} - C_{q_i}(s^{(0)})$ \; \label{line:init:bdg_update}
    Identify the next state $s^{(1)}$ for $q_i$\;  \label{line:init:nextstate}
    Compute $\Delta^{(1)}_{q_i}$ for $q_i$ by Eq.~\eqref{eq:rcu_slope} \;  \label{line:init:slope}
    $PQ.\text{push}{(q_i, s^{(1)}, \Delta^{(1)}_{q_i})}$ \; \label{line:init:push_pq}
} \label{line:init:end}
\While{$PQ$ is not empty \textbf{and} $C_{\text{bdg}} > 0$}{ \label{line:rec:start}
    $(q_j, s^{(t)}, \Delta^{(t)}_{q_j}) \gets PQ.\text{pop}()$ \; \label{line:rec:pop_pq}
    \If{$C_{bdg} - C_{q_j}(s^{(t)}) + C_{q_j}(s^{(t-1)}) < 0$ }{
        \textbf{continue} \;
    }
    $\mathcal{S}(q_j) \gets s^{(t)}$ \; \label{line:rec:update:start}
    $C_{bdg} \gets C_{bdg} - C_{q_j}(s^{(t)}) + C_{q_j}(s^{(t-1)})$\; \label{line:rec:update_bdg}
    Identify the next state $s^{(t+1)}$ for $q_j$\;  \label{line:rec:nextstate}
    \If {$s^{(t+1)}$ \textbf{exists for} $q_j$}{ \label{line:rec:next}
        Compute $\Delta^{(t+1)}_{q_j}$ for $q_j$ by Eq.~\eqref{eq:rcu_slope} \;
        $PQ.\text{push}{(q_j, s^{(t+1)}, \Delta^{(t+1)}_{q_j})}$ \;
    } 
} \label{line:rec:end}
\Return{$\mathcal{S}$}
\end{algorithm}

Algorithm~\ref{alg:greedy_upgrade} illustrates this greedy scheduling algorithm. The algorithm takes the proxy utility model $\widehat{u}_{i, k, b}$, the query set $\mathcal{Q}$ to be processed,  and the total budget $C_{bdg}$ as input, and outputs the query-to-state mapping $\mathcal{S}: q \mapsto (m_k, b)$.
First, we assign the state $(m_1, b^\text{effect}_1)$ as the initial state $s^{(0)}$ for every query in $\mathcal{Q}$ (line \ref{line:init:start}-\ref{line:init:end}), where $m_1$ is the cheapest model in the pool and $b^\text{effect}_1$ is the effective batch size of $m_1$, as calibrated in~\cref{sec:modeling}. The budget is updated for each query-state assignment in line~\ref{line:init:bdg_update}, and we compute $\Delta^{(1)}_{q_i}$ for each query $q_i$ via Eq.~\eqref{eq:rcu_slope}, which is the priority of $q_i$ for state upgrading (line~\ref{line:init:slope}). 
In line~\ref{line:init:push_pq}, we push the triplet consisting of the query, its next state $s^{(1)}$, and the priority $\Delta^{(1)}_{q_i}$ into a priority queue ${PQ}$. 
Subsequently, while the budget remains, the query $q_j$ with the largest $\Delta^{(t)}_{q_j}$ is popped from the priority queue (line~\ref{line:rec:pop_pq}), as the candidate for state upgrading along the Pareto frontier. 
As $s^{(t)}$ is an upgrade of $s^{(t-1)}$, its cost $C_{q_j}(s^{(t)})$ must be larger than $C_{q_j}(s^{(t-1)})$. The upgrade is committed only if the remaining budget is affordable for the incremental cost (line \ref{line:rec:update:start}-\ref{line:rec:end}). Here, the state of $q_j$ is updated to $s^{(t)}$ in line~\ref{line:rec:update:start} and the remaining budget is deducted in line~\ref{line:rec:update_bdg}. 
In addition, the algorithm identifies the potential next state $s^{(t+1)}$ of $q_j$ on its Pareto frontier (line~\ref{line:rec:nextstate}) and computes its cost at this state $C_{q_j}(s^{(t+1)})$.
The algorithm computes $\Delta^{(t+1)}_{q_i}$ for $q_i$ and pushes the next state triplet of $q_j$  into the priority queue (line~\ref{line:rec:next}-\ref{line:rec:end}).
This process continues until no further upgrades are available for any query or the budget is exhausted. 

\begin{example} (\emph{Running Example of Algorithm~\ref{alg:greedy_upgrade}})
Tables~\ref{tab:pareto_frontier} and~\ref{tab:pq_snapshots} present a running example of Algorithm~\ref{alg:greedy_upgrade}. 
Table~\ref{tab:pareto_frontier} lists the Pareto frontier for each query in the cost–utility space ordered by ascending state cost, while Table \ref{tab:pq_snapshots} tracks snapshots of the priority queue $PQ$ and the remaining budget $C_{bdg}$ across four successive steps. 

Initially (Step 0), all queries are assigned to their initial state $(m_1, 4)$. Given a total budget of 100, the algorithm deducts the aggregate initial cost of all queries, i.e., $\sum_{i = 1}^{6} C_{q_i} ((m_1, 4)) =9.8+10.1+9.9+10.2+10.4+10.3=60.7$, leaving a remaining budget of $C_{bdg}= 39.3$. The algorithm initializes $PQ$ by enqueuing every query $q_i$, its next candidate state $s^{(1)}$ together with the corresponding $\Delta_{q_i}^{(1)}$. The top element in $PQ$ is highlighted in blue. With $C_{bdg}>0$, the algorithm iteratively upgrades states. At Step 1, $q_1$ with the highest $\Delta_{q_i}^{(1)}$ is popped from the queue. Upgrading it from $(m_1, 4)$ to $(m_1, 2)$ incurs an additional cost of $10.7-9.8 = 0.9$. Since $C_{bdg} = 39.3 > 0.9$, the upgrade is committed and $C_{bdg}$ is updated to $39.3 - 0.9 = 38.4$. The next candidate state of $q_1$, $s^{(2)} = (m_1, 1)$, along with the corresponding $\Delta_{q_1}^{(2)} = 0.0065$ is pushed into $PQ$ as highlighted in yellow. 
In Step 2\&3, following the same logic, the algorithm upgrades $q_5$ (incremental cost 3.4), and $q_3$ (incremental cost 9.0) based on the highest $\Delta_{q_i}^{(t)}$ in $PQ$, reducing  $C_{bdg} = 38.4$ to 35.0 and 26.0 step by step. 
By continuously upgrading the most cost-effective states, \RoBatch progressively climbs the collective Pareto frontiers of all queries under the budget constraint. \eop

\end{example}

With the query-state mapping $\mathcal {S}: q_i \mapsto (m_k, b)$ determined, \RoBatch packs queries sharing the same state into batches and commits them by LLM invocations. 
\begin{theorem}
Under the routing formulation that evaluates each state by the amortized per-query cost in Eq.~\eqref{eq:per-query-cost} and the proxy utility $\widehat{u}_{i,k,b}$, pruning dominated states is lossless.
\end{theorem}
\proofsketch
Suppose there exist two states $s = (m_k, b)$ and $s' = (m_{k'}, b')$ for a query $q_i$, and that $s$ is dominated by $s'$, i.e., $s \preceq_{q_i} s'$. 
Let $\mathcal{S}$ be any feasible assignment under the formulation that assigns $q_i$ to state $s$. 
We construct another assignment $\mathcal{S}'$ by replacing the assignment of $q_i$ from $s$ to $s'$, while keeping all other query assignments unchanged.
By Definition~\ref{def:dominace}, the dominance relation $s \preceq_{q_i} s'$ implies both (1) $C_{q_i}(s) \geq C_{q_i}(s')$ and (2) $\widehat{u}_{i, k, b} \leq \widehat{u}_{i, k', b'}$. 
Consequently, replacing $q_i$'s state from $s$ with $s'$ does not decrease the total proxy utility, and does not increase the total cost under Eq.~\eqref{eq:per-query-cost}. 
Therefore, $\mathcal{S}'$ is feasible and is no worse than $\mathcal{S}$. \eop

It is worth noting that the above lossless guarantee is defined with respect to the routing formulation based on amortized per-query cost. 
This distinction does not affect the utility objective, but it may influence the exact cost usage in practice. 
The reason is that $C_{q_i}(s)$ and $C_{q_i}(s')$ in Eq.~\eqref{eq:per-query-cost} account for the system prompt cost by amortizing it over the batch sizes $b$ and $b'$, respectively. 
If the replacement leads to incomplete batches, the practical per-query system prompt cost may exceed $C_{\text{sys}}/b$ and $C_{\text{sys}}/b'$. 
In such cases, the exact total costs of $\mathcal{S}$ and $\mathcal{S}'$ become difficult to compare, as they depend on the practical sizes of the incomplete batches. 
Nevertheless, as the query workload grows, the effect of incomplete batches becomes increasingly negligible in practice.

\begin{table}[t]
    \caption{The Pareto frontier of each query $q_i$}
    \centering
    \footnotesize
    \label{tab:pareto_frontier}
    \begin{tabular}{c|l}
    \toprule
    \textbf{$q_i$} & \textbf{$s=(m_k, b)$ $(C_{q_i}(s), \widehat{u}_{i, k, b})$ $\rightarrow \cdots$ as the Pareto frontier} \\
    \midrule
    $q_1$   & $(m_1, 4) (9.8, 0.60) \rightarrow (m_1, 2) (10.7, 0.65) \rightarrow (m_1, 1) (13.8, 0.67)$  \\
    \midrule
    $q_2$   & \makecell[l]{$(m_1,4)\ (10.1, 0.60) \rightarrow (m_1,2)\ (12.9, 0.63) \rightarrow (m_1,1)\ (16.8, 0.66)$ \\
        $\rightarrow (m_2,2)\ (17.0, 0.67) \rightarrow (m_2,1)\ (19.2, 0.69)$} \\
    \midrule
    $q_3$ &  \makecell[l]{$(m_1,4)\ (9.9, 0.59) \rightarrow (m_3,4)\ (18.9, 0.69)  \rightarrow (m_3,1)\ (23.9, 0.72)$ \\}  \\
    \midrule
    $q_4$ &  \makecell[l]{$(m_1,4)\ (10.2, 0.60) \rightarrow (m_2,4)\ (14.5, 0.63) \rightarrow (m_2,2)\ (15.1, 0.65)$ \\ $\rightarrow (m_2,1)\ (19.5, 0.68)$} \\
    \midrule
    $q_5$ &  \makecell[l]{$(m_1,4)\ (10.4, 0.61)  \rightarrow (m_2,4)\ (13.8, 0.66)\rightarrow (m_2,2)\ (14.9, 0.67) $ \\            $\rightarrow (m_3,1)\ (20.2, 0.71)$} \\
    \midrule
    $q_6$ &  \makecell[l]{$(m_1,4)\ (10.3, 0.61) \rightarrow (m_1,2)\ (13.0, 0.64) \rightarrow (m_2,2)\ (14.8, 0.66) $ \\ $\rightarrow (m_2,1)\ (19.0, 0.69) \rightarrow (m_3,1)\ (24.0, 0.72)$} \\
    
    \bottomrule
    \end{tabular}
\end{table}

\begin{table}[t]
    \caption{Four snapshots of $PQ$: each cell is $(q_i, s_i^{(t)}, \Delta_{q_i}^{(t)})$. }
    \centering
    \scriptsize
    \label{tab:pq_snapshots}
    \setlength{\tabcolsep}{3pt}
    \renewcommand{\arraystretch}{1.05}
    \begin{tabular}{p{0.23\columnwidth} | p{0.23\columnwidth} |p{0.23\columnwidth} |p{0.23\columnwidth}}
    
    \toprule
    \textbf{Step 0} & \textbf{Step 1} & \textbf{Step 2} & \textbf{Step 3}  \\
    
    \midrule
     \textbf{$C_{bdg}=$39.3} & \textbf{$C_{bdg}=$38.4}& \textbf{$C_{bdg}=$35}&\textbf{$C_{bdg}=$26}\\    
     \midrule

    \cellcolor{LightCyan}{$(q_1,(m_1,2),0.0556)$} & \cellcolor{LightCyan}{$(q_5,(m_2,4),0.0147)$} & \cellcolor{LightCyan}{$(q_3,(m_3,4),0.0111)$} & \cellcolor{LightCyan}{$(q_6,(m_1,2),0.0111)$} \\
    $(q_5,(m_2,4),0.0147)$ & $(q_3,(m_3,4),0.0111)$ & $(q_6,(m_1,2),0.0111)$ & $(q_2,(m_1,2),0.0107)$ \\
    $(q_3,(m_3,4),0.0111)$ & $(q_6,(m_1,2),0.0111)$ & $(q_2,(m_1,2),0.0107)$ & $(q_5,(m_2,2),0.0091)$ \\
    $(q_6,(m_1,2),0.0111)$ & $(q_2,(m_1,2),0.0107)$ & \cellcolor{LightYellow}{$(q_5,(m_2,2),0.0091)$} & $(q_4,(m_2,4),0.0070)$ \\
    $(q_2,(m_1,2),0.0107)$ & $(q_4,(m_2,4),0.0070)$ & $(q_4,(m_2,4),0.0070)$ & $(q_1,(m_1,1),0.0065)$ \\
    $(q_4,(m_2,4),0.0070)$ & \cellcolor{LightYellow}{$(q_1,(m_1,1),0.0065)$} & $(q_1,(m_1,1),0.0065)$ & \cellcolor{LightYellow}{$(q_3,(m_3,1),0.0060)$} \\


    \bottomrule
    \end{tabular}
\end{table}

\stitle{Complexity Analysis.}
\label{routing:complexity}
The routing stage has an overall time complexity of $\mathcal{O}(|\mathcal{Q}| (K d + \widetilde{\mathcal{B}} \log \widetilde{\mathcal{B}} + T \log |\mathcal{Q}|))$, where $\widetilde{\mathcal{B}} =\sum_{k=1}^K |\mathcal{B}_k|$ denotes the total number of candidate states across all models, $T$ denotes the maximum number of states in the Pareto frontier across all queries, and $d$ is  query embedding dimensionality. This complexity comprises three parts. First, the proxy utility model predicts the utilities of the $\widetilde{\mathcal{B}}$ states for the $|\mathcal{Q}|$ queries, requiring $\mathcal{O}(|\mathcal{Q}| (K d + \widetilde{\mathcal{B}}))$ complexity. Second, identifying the Pareto frontier necessitates sorting the non-dominated states for each query, which contributes $\mathcal{O}(|\mathcal{Q}|\widetilde{\mathcal{B}} \log \widetilde{\mathcal{B}})$ at most. 
Third, the greedy scheduling takes a complexity of $\mathcal{O}(|\mathcal{Q}| T \log |\mathcal{Q}|)$, because each of the $|\mathcal{Q}|$ queries can be enqueued and dequeued into the priority queue at most $T$ times, with each push or pop operation costing $\mathcal{O}(\log |\mathcal{Q}|)$.

\section{Experimental Study}
\label{sec:exp}

In this section, we present the experimental setup in \cref{sec:exp:setup} and evaluate our approach in the following facets: 
\ding{172} Compare with the baseline LLM routing and batching approaches (\cref{sec:exp:performance}); 
\ding{173} Conduct the ablation studies for \RoBatch (\cref{sec:exp:ablation});
\ding{174} Study the robustness of \RoBatch to key hyper-parameters and design choices (\cref{sec:exp:sensitivity}); 
\ding{175} Compare the scalability and scheduling overhead of \RoBatch with the baselines (\cref{sec:exp:scalability}).

\subsection{Experimental Setup}
\label{sec:exp:setup}

\subsubsection{Datasets}
To evaluate the effectiveness and efficiency of \RoBatch, we conduct experimental studies on six widely used NLP benchmarks from four task categories as follows: 
\begin{itemize}[leftmargin =*]
\item \textbf{Mathematical Reasoning.} \GSM~\cite{DBLP:journals/corr/abs-2110-14168} is a collection of 8,000 high-quality grade school mathematical word problems.
\item \textbf{General Knowledge QA.} \MMLU~\cite{DBLP:conf/iclr/HendrycksBBZMSS21} is a comprehensive benchmark covering 57 subjects across STEM, the humanities, and the social sciences.
\item \textbf{Commonsense and Semantic Reasoning.} \SNLI (Stanford Natural Language Inference)~\cite{DBLP:conf/emnlp/BowmanAPM15} evaluates a model’s ability to determine the logical relationship between a pair of sentences, i.e., entailment, neutrality, or contradiction. \MRPC (Microsoft Research Paraphrase Corpus)~\cite{DBLP:conf/acl-iwp/DolanB05} consists of sentence pairs annotated for semantic equivalence, and is used to assess fine-grained semantic understanding.
\item \textbf{Text Classification.} \AGNews~\cite{DBLP:conf/nips/ZhangZL15} is a topic classification benchmark in which articles are categorized into four classes: World, Sports, Business, and Science/Technology. \IMDB~\cite{DBLP:conf/acl/MaasDPHNP11} is a binary sentiment classification dataset consisting of movie reviews and is widely used to evaluate sentiment understanding.
\end{itemize}





\subsubsection{Baselines}
We compare \RoBatch against two LLM routing and two batch prompting baselines as follows: 
\begin{itemize}[leftmargin =*]
    \item \textbf{LLM Routing.} \RouteLLM~\cite{DBLP:conf/iclr/OngAWC0GKS25} is a preference-data-driven routing framework that selects between a stronger and a weaker LLM. For each query, it predicts whether the stronger model is necessary and uses a threshold to trade off response quality against inference cost. \FrugalGPT~\cite{DBLP:journals/tmlr/ChenZ024} is a budget-aware framework for routing queries across multiple LLMs. It employs a learned utility estimator and threshold-based decision rules to determine whether a cheaper or a stronger model should be used.
    \item \textbf{Batch Prompting.} \BATCHER~\cite{DBLP:conf/icde/FanHFC00024} is originally an LLM-based entity resolution framework, but it adopts a similarity-based  (\BATCHERSIM) and a diversity-based (\BATCHERDIV) batching strategies. \BATCHERSIM groups semantically similar queries into the same batch, whereas \BATCHERDIV constructs each batch using queries drawn from different clusters to increase diversity. 
    \OBP~\cite{DBLP:journals/pvldb/JiWLXZ25} is an optimized batch-prompting framework designed to reduce LLM inference cost through adaptive query grouping. Instead of relying on fixed-size batching, it clusters related queries and dynamically refines the resulting groups to balance query affinity, context length, and accuracy. 
\end{itemize}
Since these baselines do not jointly support both routing and batching, in the experiments, we adapt these methods to ensure a fair comparison. 
Specifically, for the LLM routing baselines, we first apply the router to assign each query to a target model and then group the queries assigned to the same model into batches of a fixed size.
For batch prompting baselines, we use the same router for non-batch prompting of our \RoBatch, built in \cref{sec:modeling}, and then construct batches for each model according to the batching strategy of the baselines.

\subsubsection{Evaluation metrics}
We use three metrics to evaluate the effectiveness and efficiency of our framework. For the utility, we report the Average (Avg.) Accuracy of the queries is the primary objective in our framework. Apart from the classification tasks, Accuracy indicates whether the LLM correctly solves the problem instance. 
For monetary cost, we report the total expenditure (in \$) incurred by each approach across the test workload.
For computational cost, we report the wall-clock time (in seconds) for batching and/or routing algorithms, excluding the latency of LLM API calls.




\subsubsection{Implementation details}

Our framework \RoBatch is implemented by PyTorch 2.10.0+cu128 with Python 3.10.19. We use the Qwen3-0.6B~\cite{DBLP:journals/corr/abs-2506-05176} as the embedding model to generate the features of 1,024 dimensions for the router models of \RoBatch, and adopt two vanilla ML models, a three-layer MLP and a KNN classifier. 
Consistent with recent work on batch prompting~\cite{DBLP:conf/emnlp/ChengK023}, we adopt a budget-conscious sampling strategy to limit the high cost of exhaustive API evaluation. Specifically, for each benchmark, we partition the dataset into 2,048 training instances, 512 validation instances, and 1,024 test instances. The training set is used primarily for the modeling stage, while the validation set is used for hyperparameter tuning. Final results are reported on the unseen test set to assess generalization.
We perform a grid search over key hyperparameters, including the number of hidden layers, learning rate, and dropout ratio of the MLP, and select the configuration that achieves the highest Accuracy. 
In addition, \RoBatch uses a threshold $\epsilon =1\%$ (in Eq.~\eqref{eq:batchsize:bound}) and uses the $k$-center greedy algorithm to find a coreset with size 256 from the 2,048 training instances. All batch size $b_k \in \mathcal{B}_k$ used are multiples of 4.
For invoking LLM instances, we use the native Qwen API for the Qwen3 family~\cite{DBLP:journals/corr/abs-2505-09388} (Qwen3-4B, Qwen3-14B and Qwen3-32B) and the OpenRouter API for the Gemma3 family~\cite{DBLP:journals/corr/abs-2503-19786} (Gemma3-4B, Gemma3-12B and Gemma3-27B), ensuring a standardized request-response workflow across models.
In addition, we disable thinking mode and use structured input and output via JSON format to prompt the LLM to generate exactly $b$ outputs.
All experiments are conducted on a server equipped with NVIDIA RTX 4090 GPU with 24G memory to process embedding generation and router training, and the scheduling algorithm is running on the same server with Intel(R) Xeon(R) Silver 4316 CPU and 512G memory. 




\subsection{Overall Performance}
\label{sec:exp:performance}

\begin{figure*}[ht]
    \captionsetup[subfigure]{labelformat=empty}
	\includegraphics[width=1.5\columnwidth]{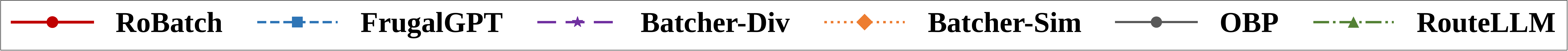}
    \setlength{\tabcolsep}{0pt}
     \begin{tabular}[h]{@{}cccccc@{}}
        
        \subfigure[\GSM] {
				\includegraphics[width=0.33\columnwidth]{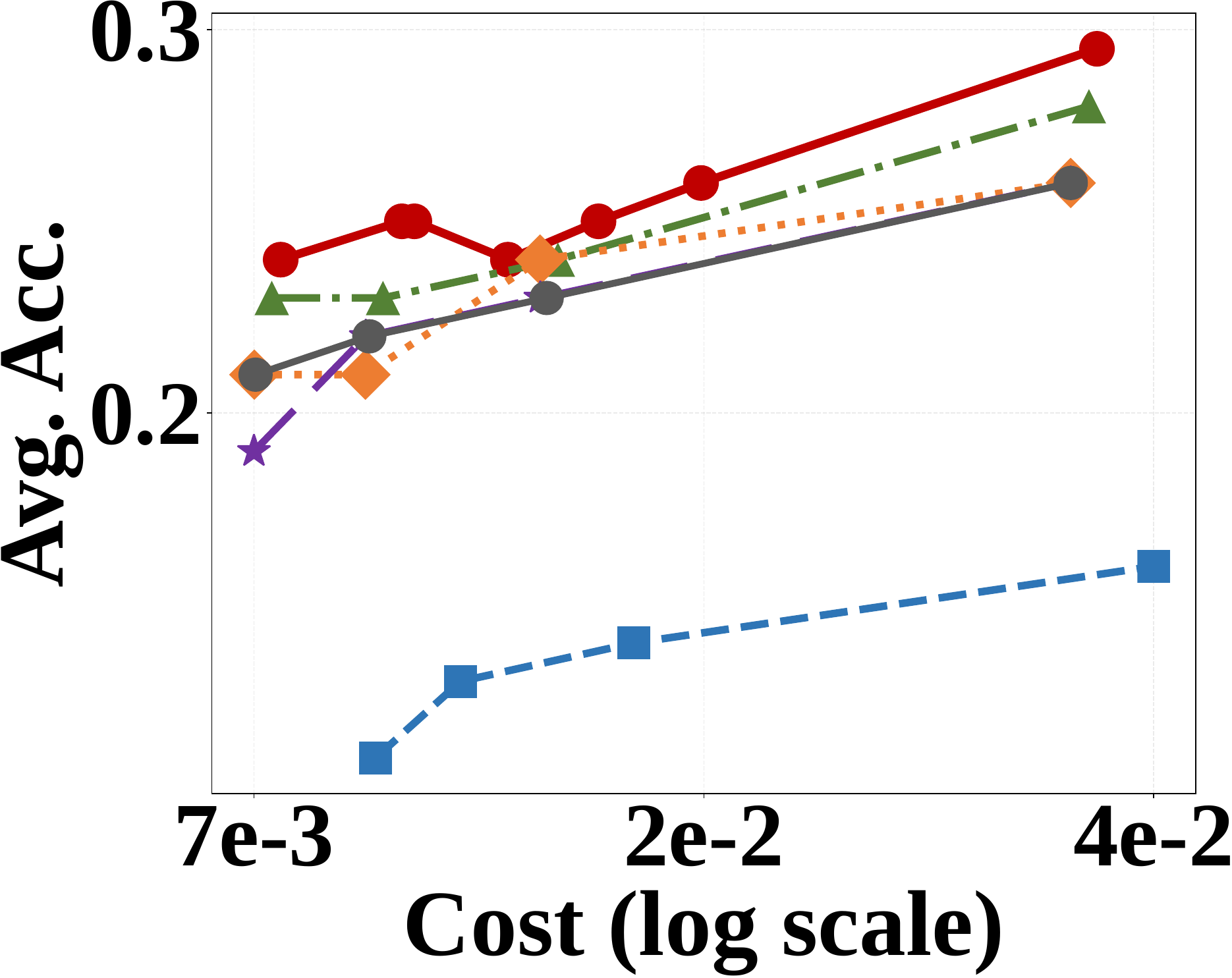}
			\label{fig:exp:gemma:gsm8k}
		} &
        \subfigure[MMLU] {
				\includegraphics[width=0.33\columnwidth]{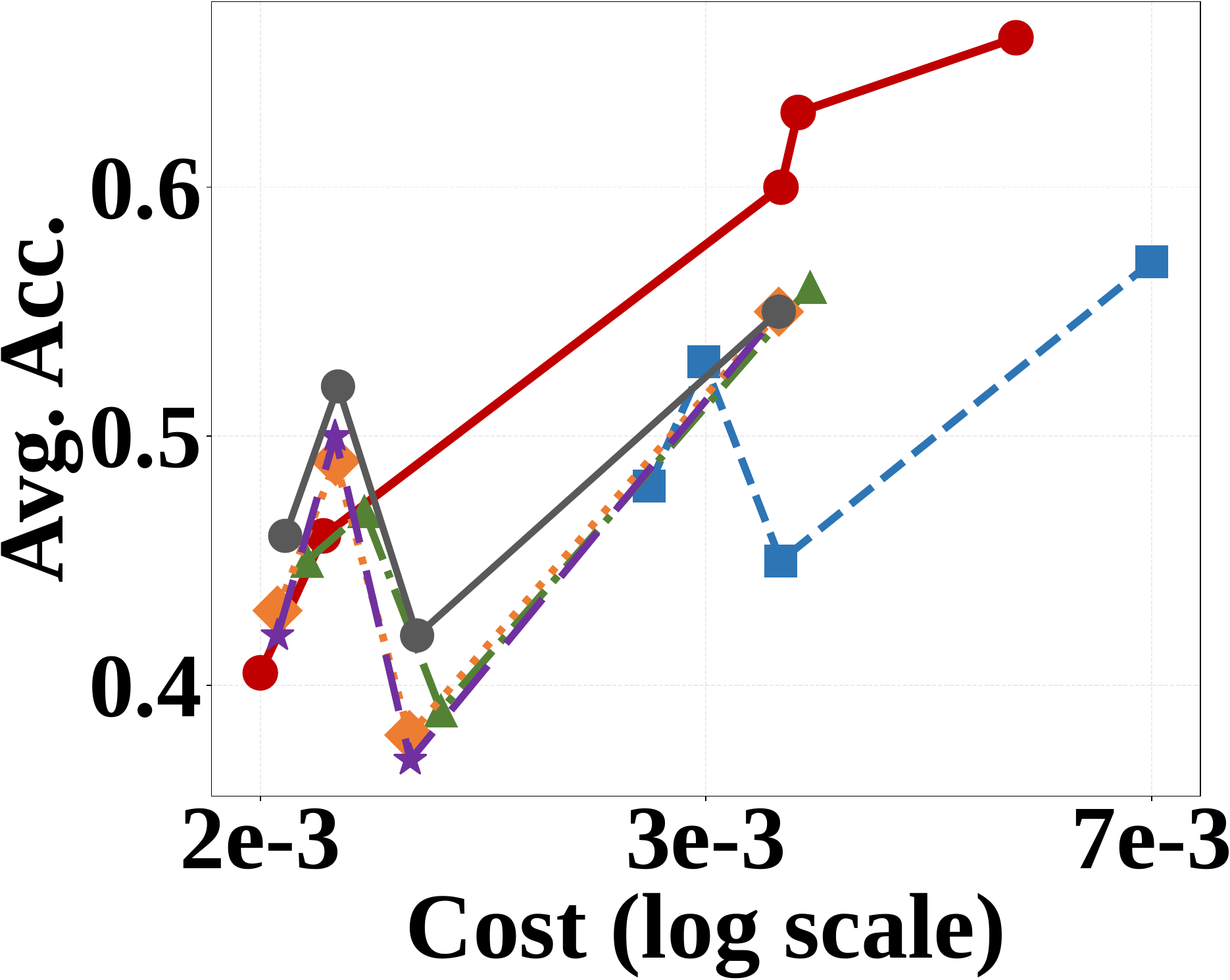}
			\label{fig:exp:gemma:mmlu}
		} &
        \subfigure[SNLI] {
				\includegraphics[width=0.33\columnwidth]{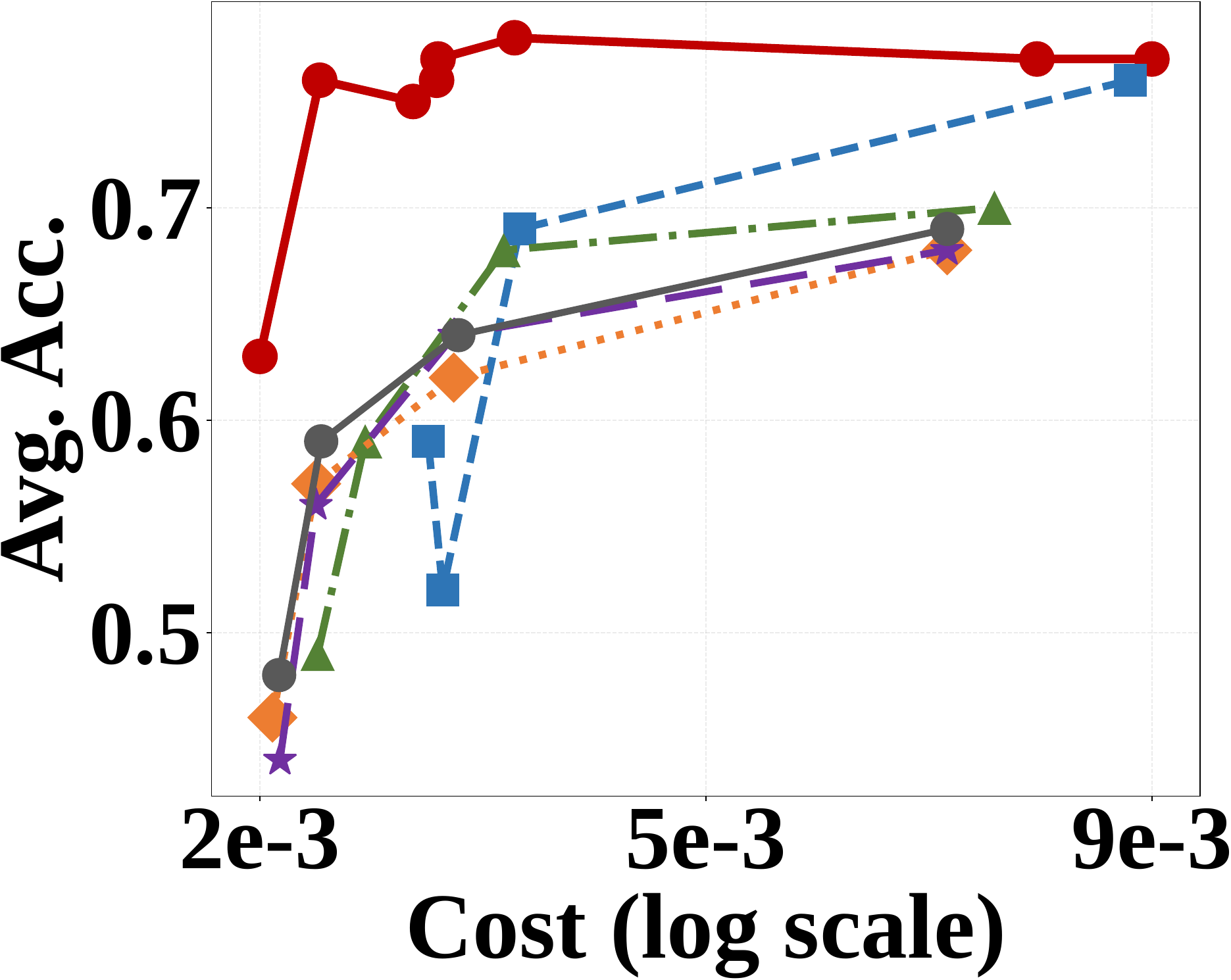}
			\label{fig:exp:gemma:snli}
		} &
        \subfigure[\MRPC] {
				\includegraphics[width=0.33\columnwidth]{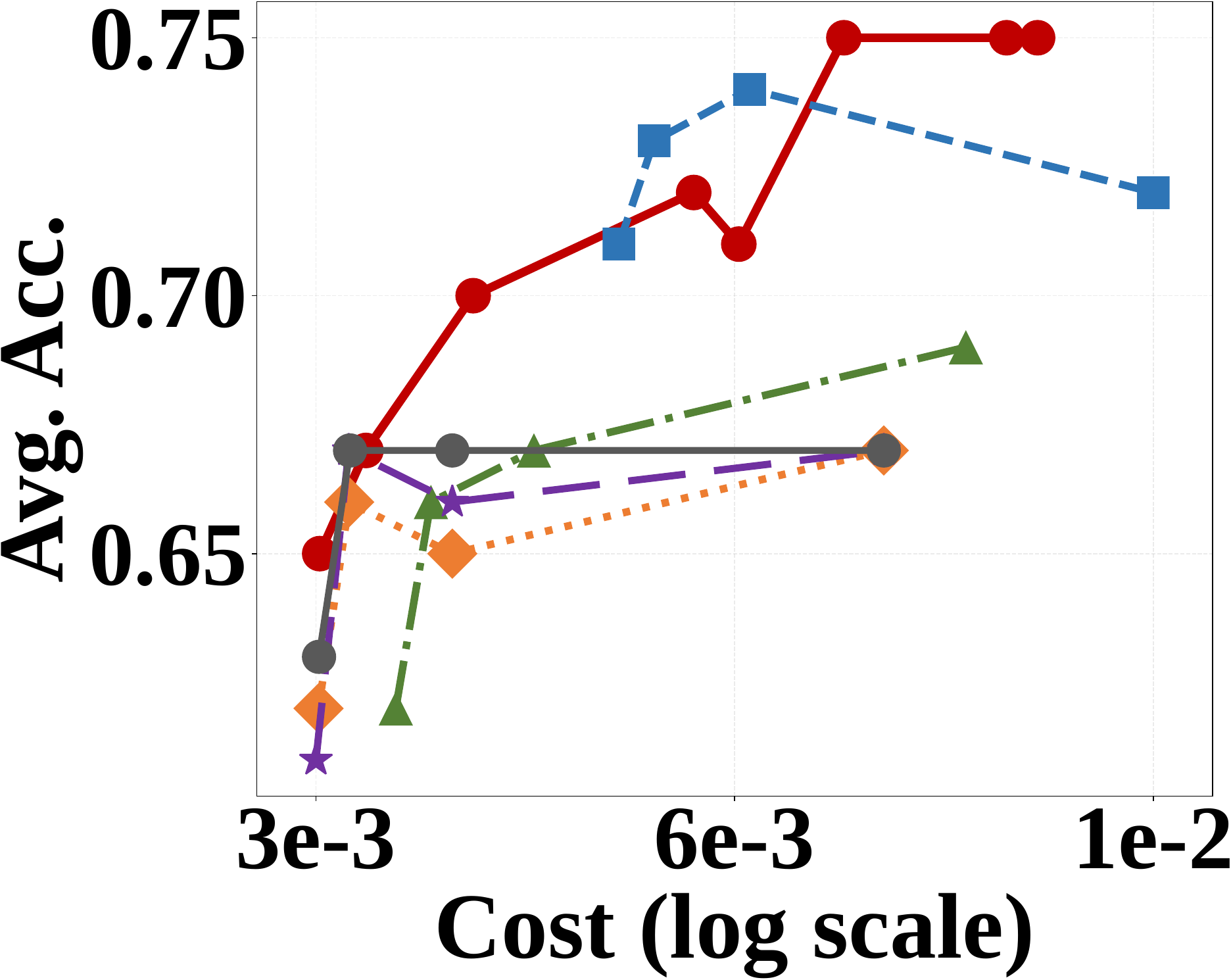}
			\label{fig:exp:gemma:mrpc}
		} &
        \subfigure[\AGNews] {
				\includegraphics[width=0.33\columnwidth]{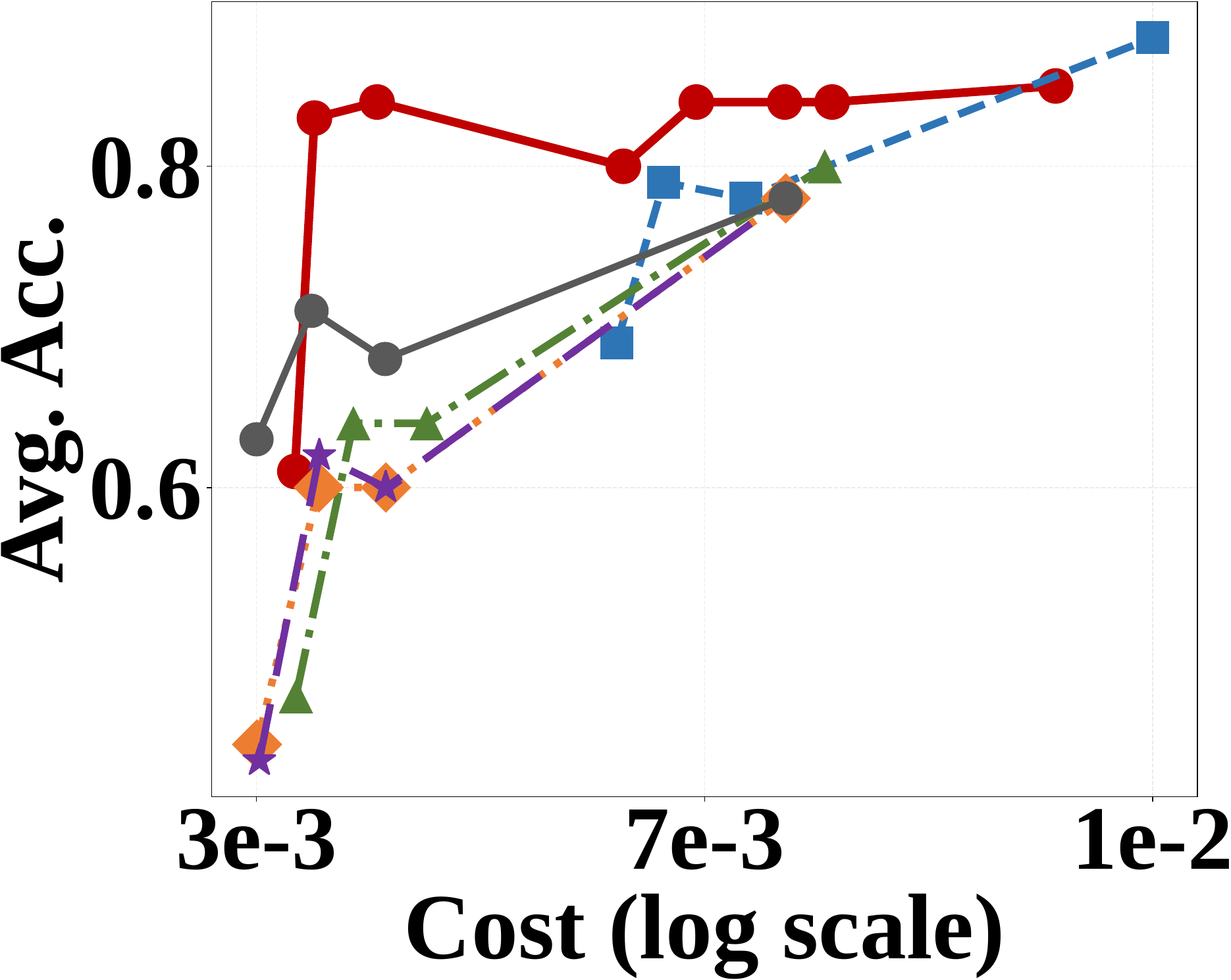}
			\label{fig:exp:gemma:agnews}
		} &

        \subfigure[IMDB] {
				\includegraphics[width=0.33\columnwidth]{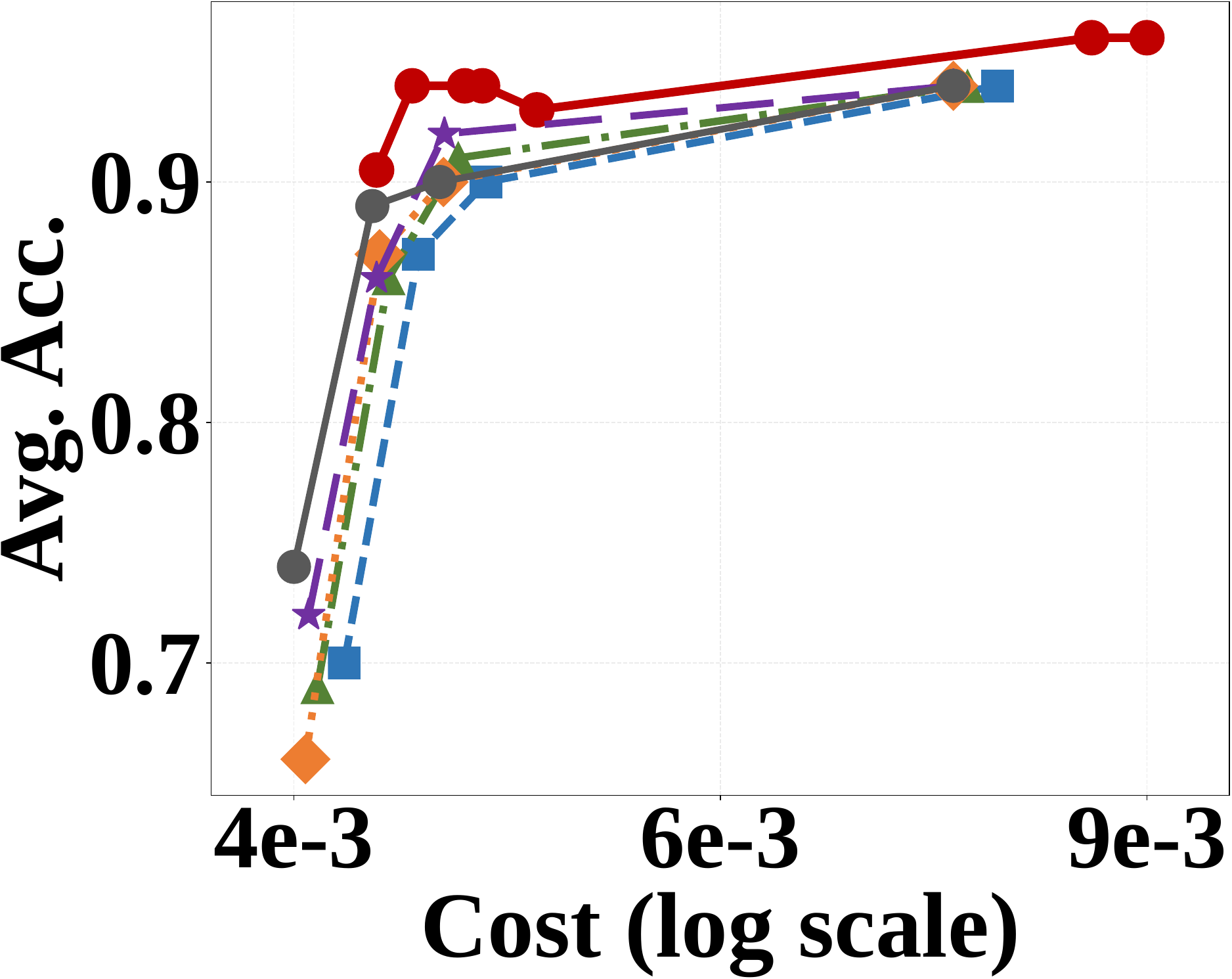}
			\label{fig:exp:gemma:imdb}
		}  \\

        \multicolumn{6}{c}{Gemma3 (Gemma3-4B, Gemma3-12B, Gemma3-27B)} \\

        \subfigure[\GSM] {
				\includegraphics[width=0.33\columnwidth]{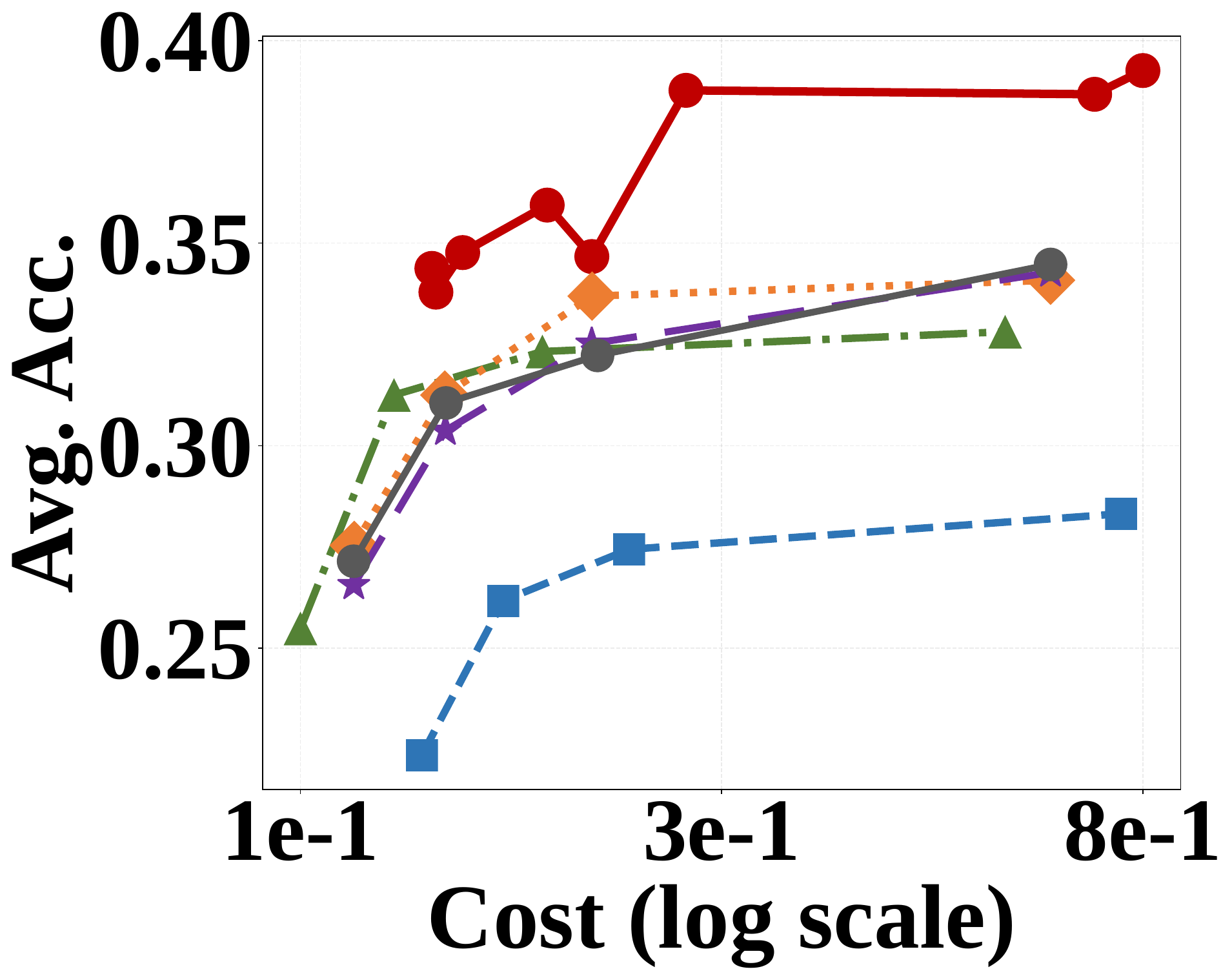}
			\label{fig:exp:qwen:gsm8k}
		} &
        \subfigure[MMLU] {
				\includegraphics[width=0.33\columnwidth]{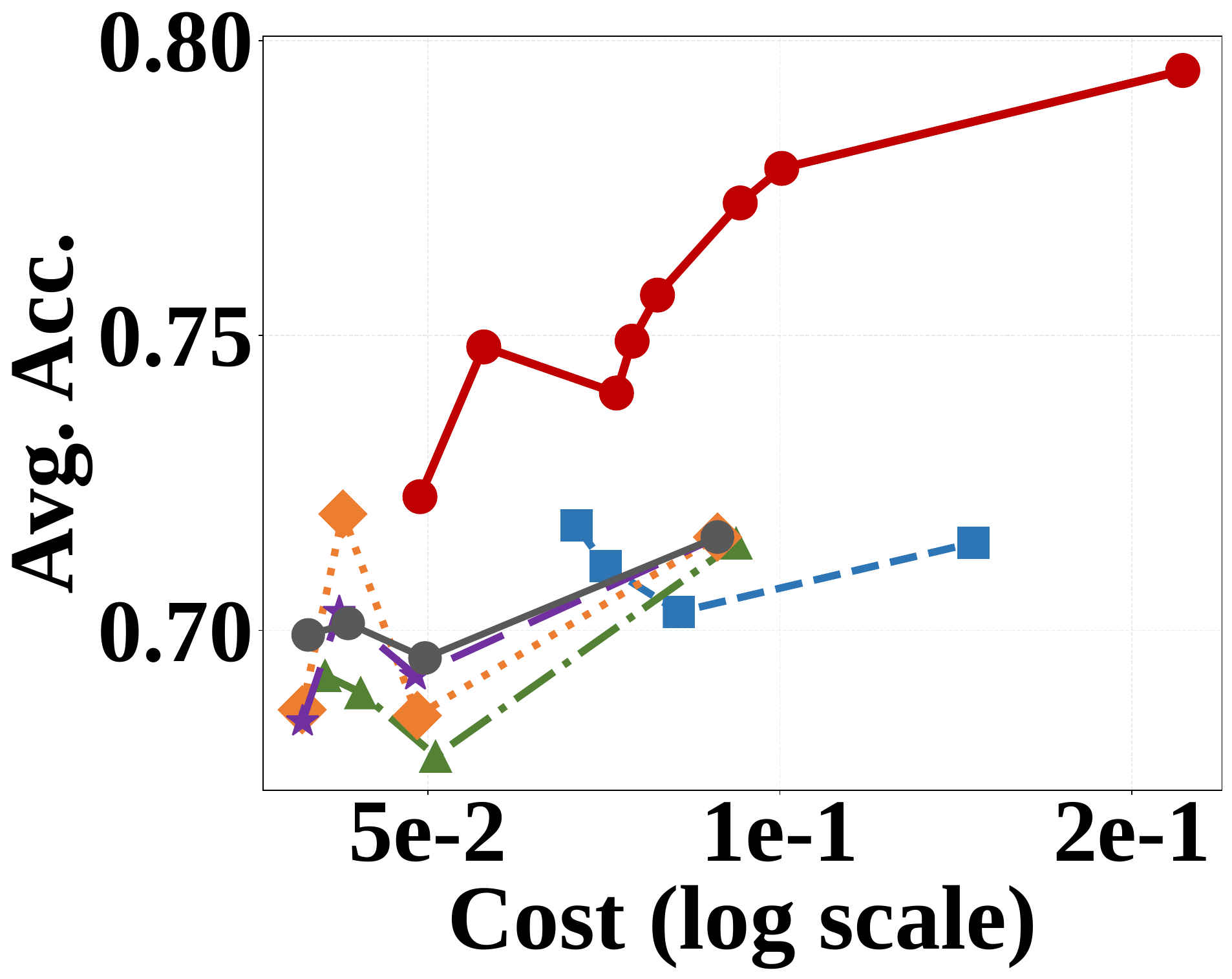}
			\label{fig:exp:qwen:mmlu}
		} &
        \subfigure[SNLI] {
				\includegraphics[width=0.33\columnwidth]{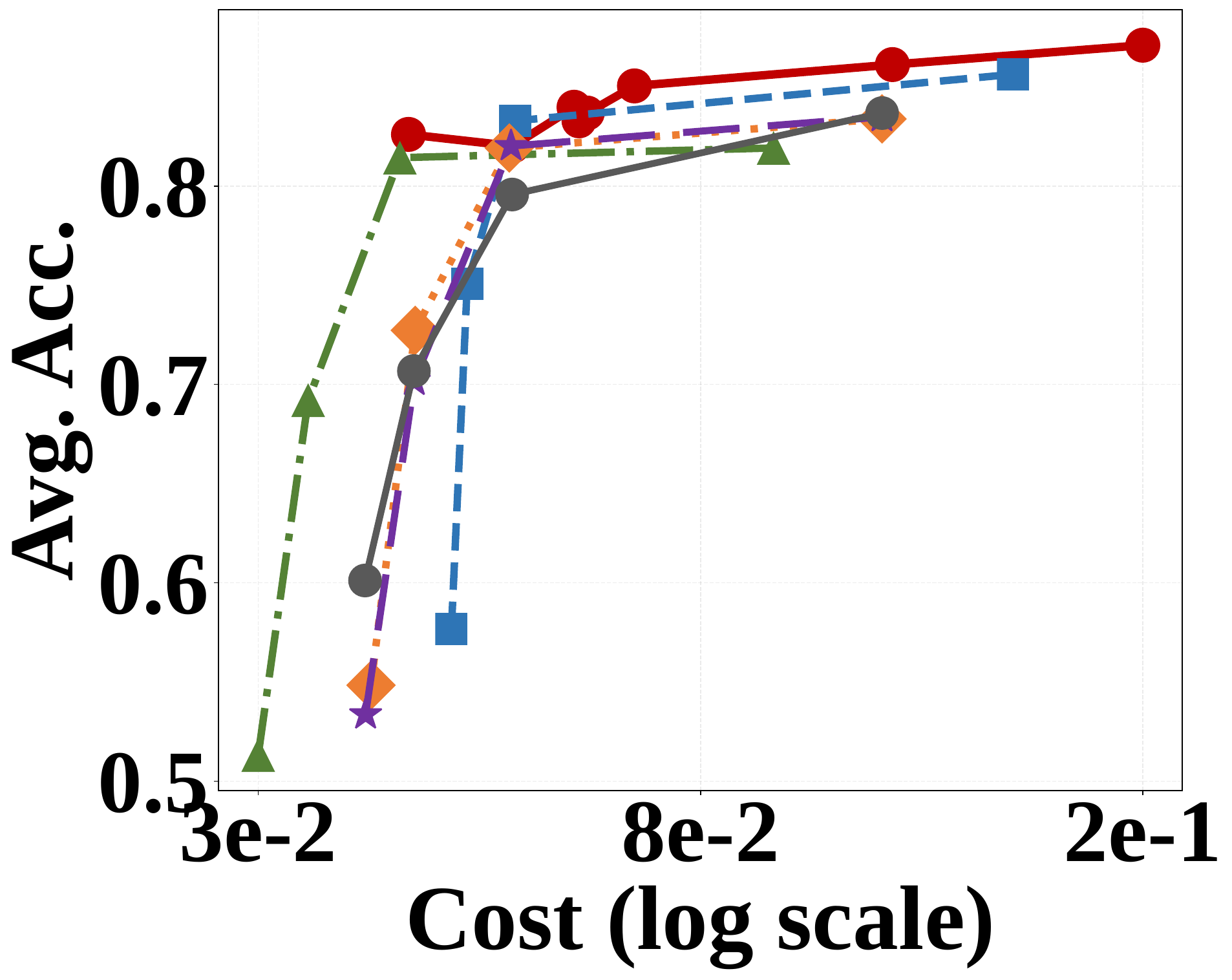}
			\label{fig:exp:qwen:snli}
		} &
        \subfigure[\MRPC] {
				\includegraphics[width=0.33\columnwidth]{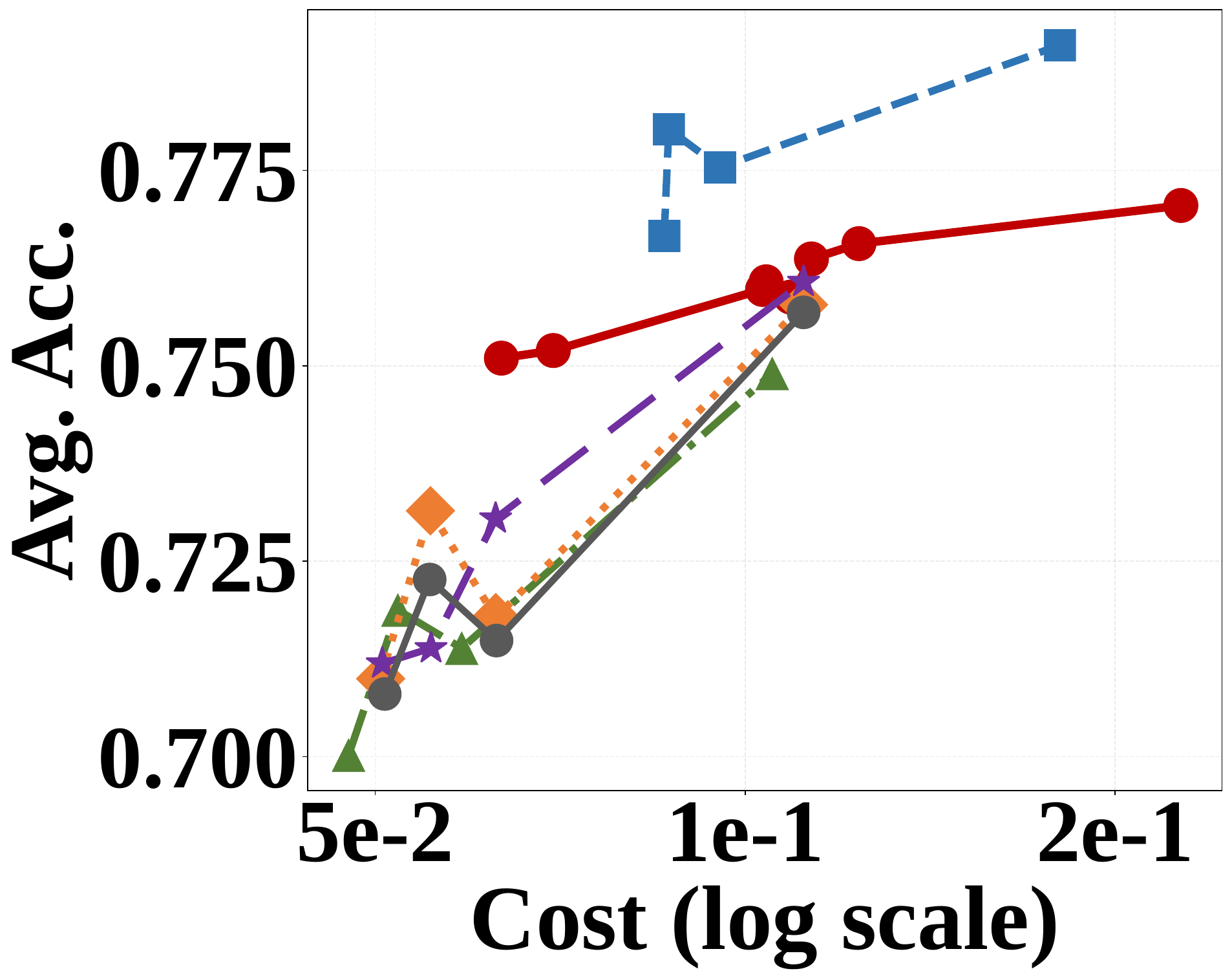}
			\label{fig:exp:qwen:mrpc}
		} &
        \subfigure[\AGNews] {
				\includegraphics[width=0.33\columnwidth]{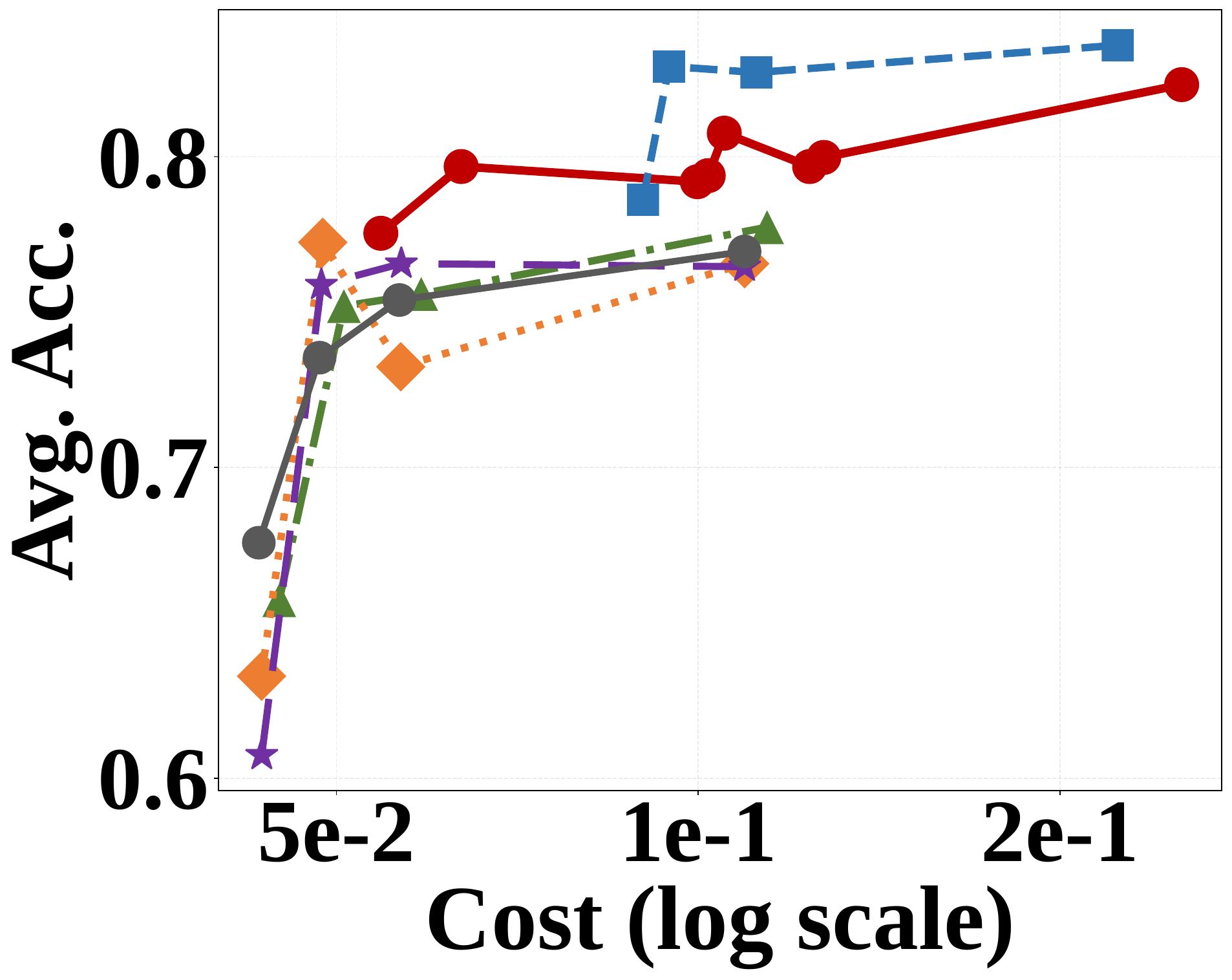}
			\label{fig:exp:qwen:agnews}
		} &

        \subfigure[IMDB] {
				\includegraphics[width=0.33\columnwidth]{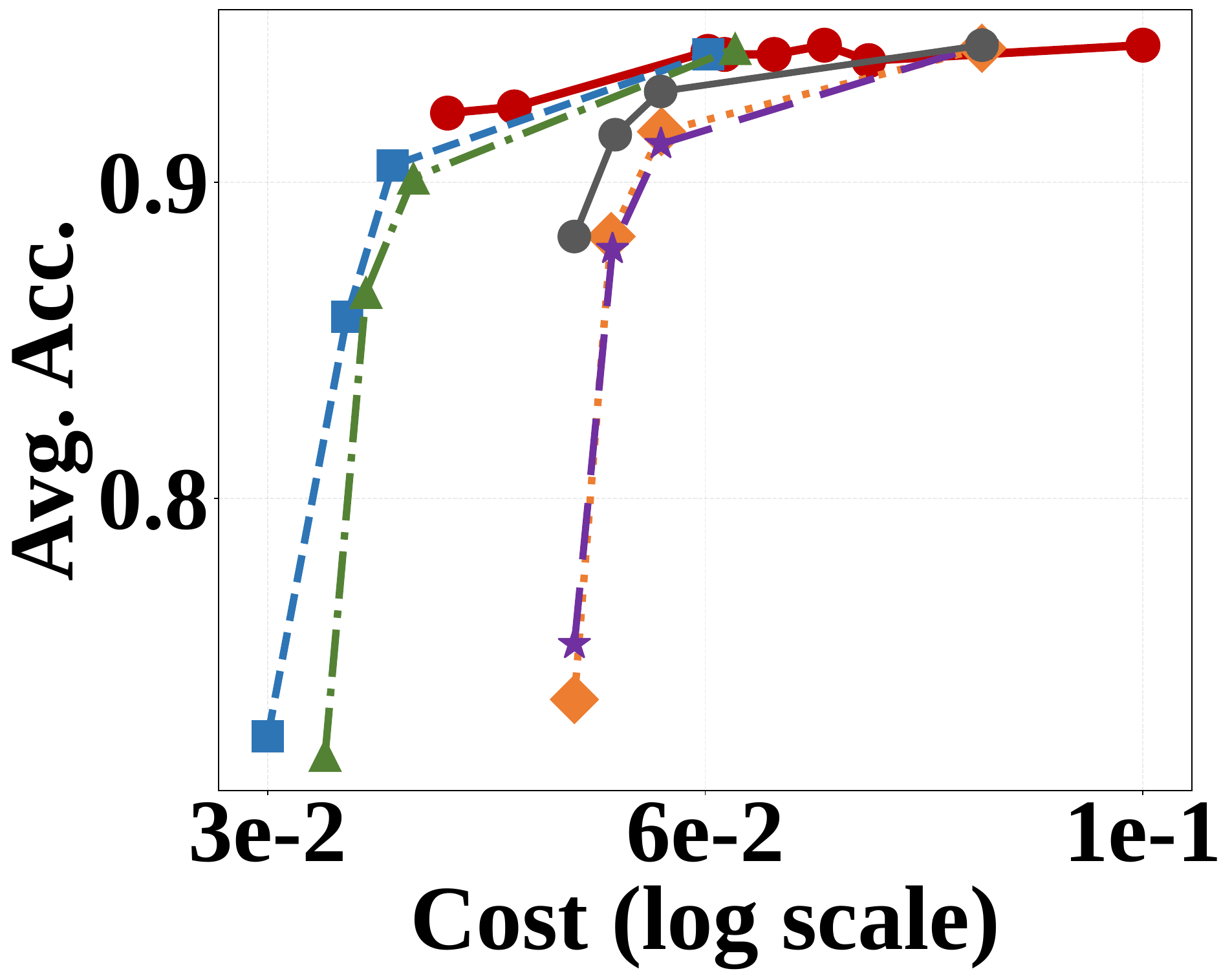}
			\label{fig:exp:qwen:imdb}
		}  \\

        \multicolumn{6}{c}{Qwen3 (Qwen3-4B, Qwen3-14B and Qwen3-32B)} \\
        
	\end{tabular}
 
	\caption{Overall Cost–Accuracy Trade-Off on Six Benchmarks with the Gemma3 and Qwen3 Model Families}
    \label{fig:exp:main}
\end{figure*}

We compare the overall performance of \RoBatch against the routing and batching baselines. 
For baseline methods that require a fixed batch size, we instantiate the batch size $b \in \{16, 8, 4, 1\}$, resulting in four corresponding budget levels. For each budget level, the real cost usages of the baselines are different because of their different routing strategies, and we take the maximum and minimum costs of the baselines as two budgets for \RoBatch. 
Fig.~\ref{fig:exp:main} presents the overall cost-accuracy landscape across six benchmarks under
both the Gemma3 and Qwen3 model families. Here, the cost on the x-axis is the actual spent cost rather than the budget. 

Overall, \RoBatch delivers the most favorable cost–accuracy trade-off across a broad range of workloads and budget regimes. In most figures, the \RoBatch curve dominates the cost-accuracy trade-off, indicating that it achieves higher accuracy at the same monetary cost, or equivalently, reaches the same accuracy with lower cost than the baselines. These results suggest that the advantage of \RoBatch comes from its ability to optimize model selection and batch prompting decisions in a unified manner rather than treating them as two independent steps.

The advantage of \RoBatch is particularly remarkable in reasoning and knowledge-intensive tasks, such as \GSM, \MMLU, and \SNLI, where \RoBatch clearly achieves a better cost–performance trade-off than the baselines. The improvement is especially evident on \GSM and \MMLU, indicating that the joint optimization strategy can allocate budget more effectively on difficult workloads by selecting more cost-effective assignment for each query. On \SNLI, although the absolute gap among methods narrows relative to \GSM and \MMLU, \RoBatch still lies on, or very near, the optimal frontier overall. This suggests that the benefit of \RoBatch is not confined to a single type of tasks, but generalizes well across different types of complex workloads. We also observe that the performance gap is generally larger under the Qwen3 family, while under the Gemma3 family it becomes narrower on some workloads, although the overall trend remains consistent.

In contrast, on relatively easy classification tasks such as \AGNews and \IMDB, the performance gap among methods becomes smaller. As illustrated in Fig.~\ref{fig:exp:main}(\subref{fig:exp:qwen:agnews},\subref{fig:exp:gemma:agnews},\subref{fig:exp:qwen:imdb},\subref{fig:exp:gemma:imdb}), multiple methods achieve very similar performance in the high-cost region, and their main distinction lies in how effectively they improve accuracy under low-cost regimes. Despite this, \RoBatch continues to perform favorably relative to the baselines and still achieves the best or competitive cost–accuracy frontier over most cost regimes. A similar pattern can also be observed on \MRPC, where the gain of \RoBatch is more moderate than that on the harder reasoning-oriented tasks, yet its overall curve still exhibits better economic efficiency. This indicates that when a task becomes less sensitive to model capacity and execution strategy, the benefit of joint optimization is reflected mainly in maintaining competitive accuracy at lower cost.

In summary, Fig.~\ref{fig:exp:main} reveals a consistent trend: \RoBatch more demonstrably improves the cost-accuracy frontier on hard tasks, while on easier tasks it more reliably reduces the cost needed to reach comparable Accuracy, thereby yielding a stronger overall frontier across diverse workloads.

\subsection{Ablation Studies}
\label{sec:exp:ablation}

\begin{figure}[t]
\includegraphics[width=0.9\columnwidth]{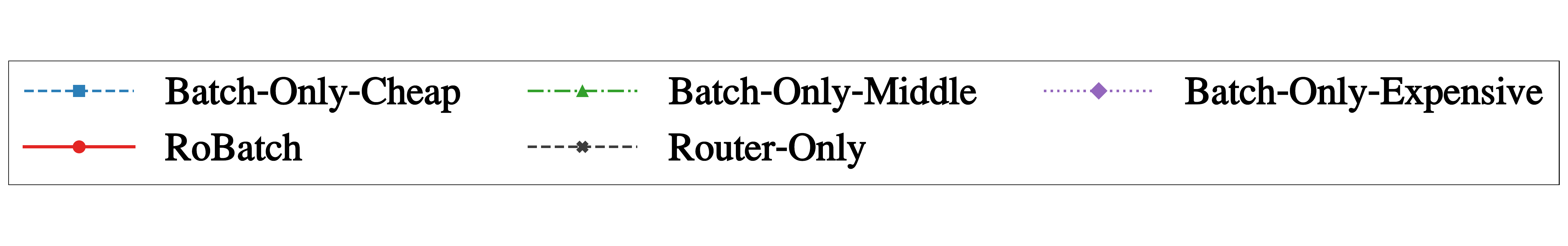}

 \begin{tabular}[h]{c}
 
        \subfigure[\GSM] {
				\includegraphics[width=0.32\columnwidth]{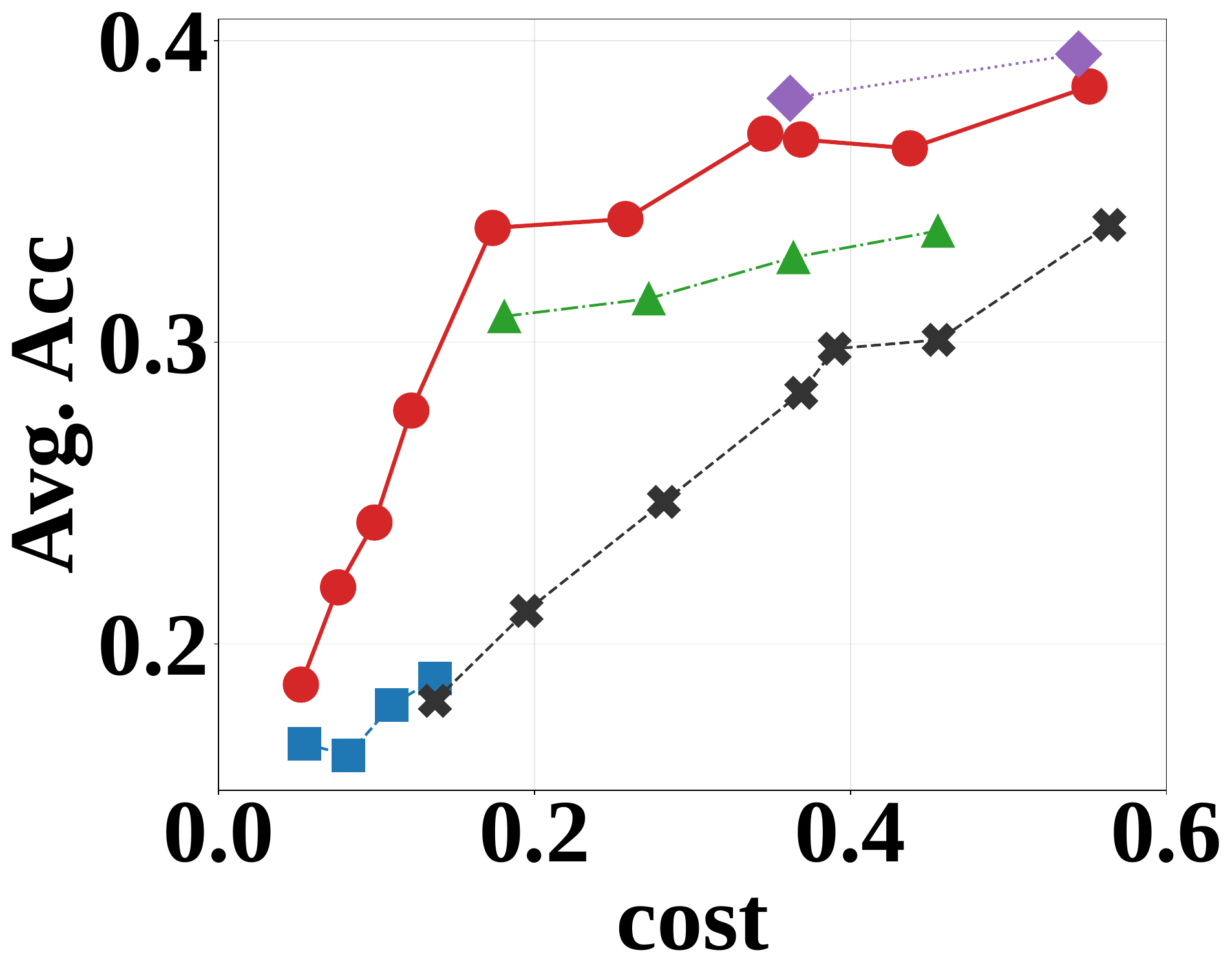}
			\label{fig:exp:ablation:gsm8k}
		}\hspace{-2ex}         
        \subfigure[\AGNews] {
				\includegraphics[width=0.32\columnwidth]{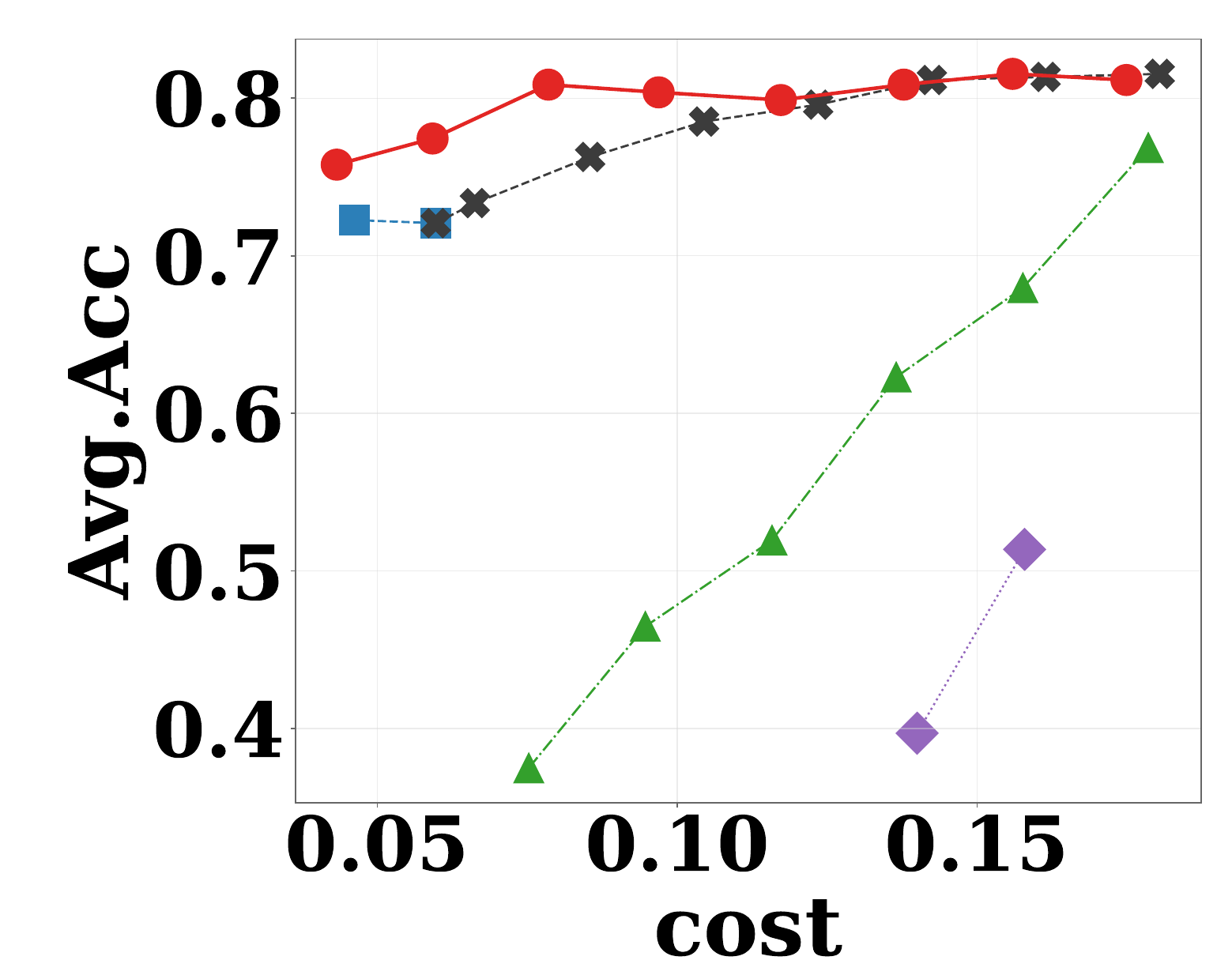}
			\label{fig:exp:ablation:ag_news}
		}  \hspace{-2ex}         
        \subfigure[\IMDB] {
				\includegraphics[width=0.32\columnwidth]{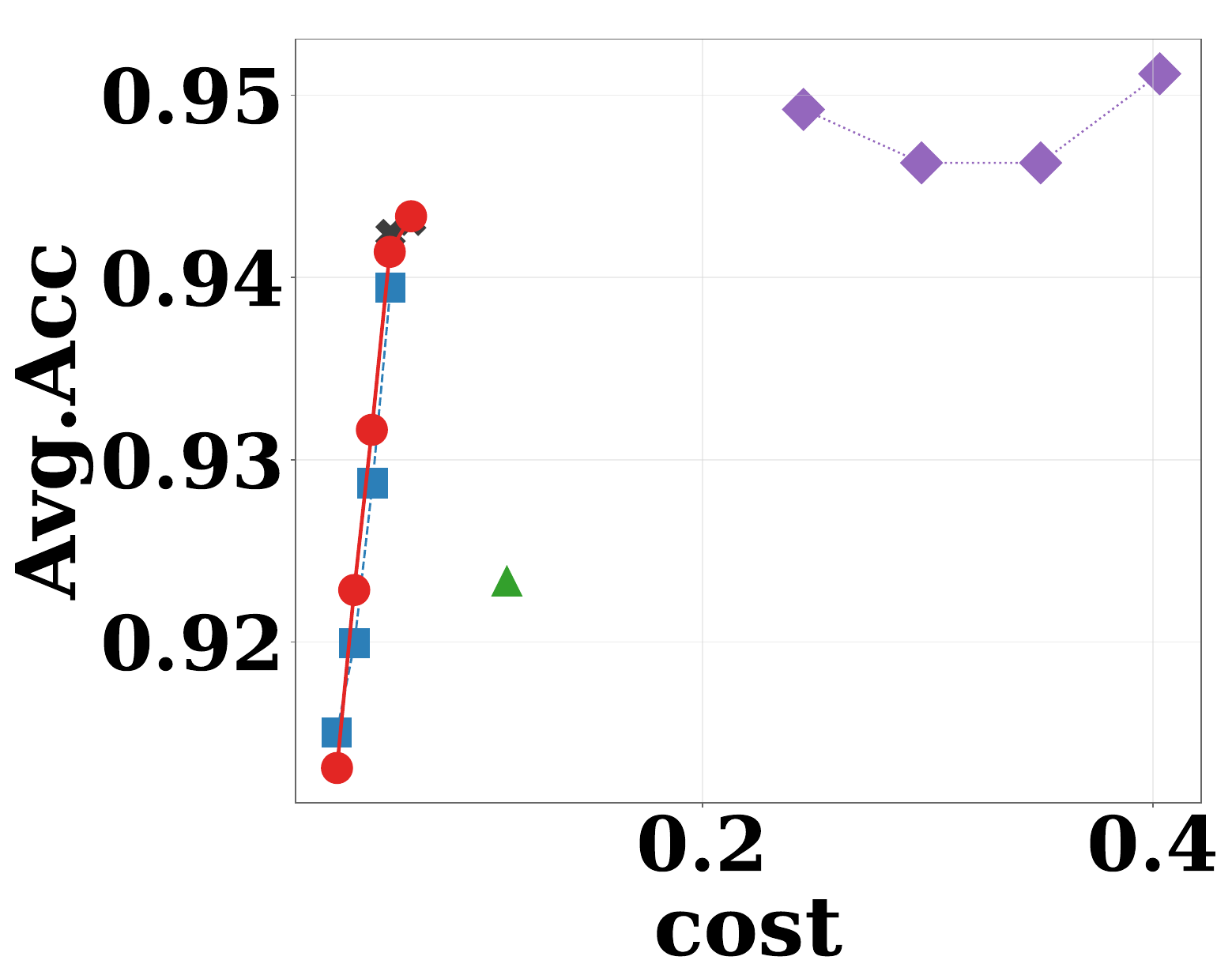}
			\label{fig:exp:ablation:imdb}
		}         

\end{tabular}

	\caption{Ablation Studies: Comparison of \RoBatch with Router-Only and Batch-Only Counterparts}

    \label{fig:exp:ablation}

\end{figure}

To evaluate the individual contributions of the routing and batching components within our framework, we conduct a series of ablation experiments that decouple the joint optimization strategy into two ablated variants: a Router-Only approach and a Batch-Only approach. Router-Only fixes $b=1$ for all queries, reducing the state space to the $K$ candidate models in $\mathcal{M}$ and isolating model selection via multi-label scoring without benefiting from prompt amortization. Batch-Only assigns all queries to one fixed model. Accordingly, we instantiate three Batch-Only variants in the experiments, corresponding to the cheap, middle, and expensive models, respectively. Within each variant, the greedy scheduling process is restricted to transitions over the fixed model $m_k$ with batch sizes in the model-specific valid space $\mathcal{B}_k$.

As shown in Fig.~\ref{fig:exp:ablation}, the full \RoBatch consistently yields a superior cost-accuracy frontier than either ablated variant across the three tasks. In particular, the improvement is most evident in the low and medium-cost regions, where \RoBatch achieves clearly higher accuracy at comparable cost. This trend is particularly evident in the more challenging reasoning workloads \GSM, where neither Router-Only nor Batch-Only is sufficient to recover the same frontier. On the easier classification workloads \AGNews and \IMDB, the gap narrows as the cost increases, and the curves of \RoBatch and Router-Only gradually converge. For example, on \AGNews, both methods eventually approach a similar high accuracy (around 0.81), while on \IMDB most approaches already approach the performance ceiling (around 0.94-0.95), leaving less room for additional gains.

 Router-Only can improve model assignment at the query level, but each LLM invocation processes only one system prompt and one query and therefore cannot fully exploit the cost savings enabled by prompt amortization. In contrast, Batch-Only is constrained by the capability and pricing of the single model it uses. Under very tight budgets, increasing the batch size may still be insufficient to meet the budget constraint if the model itself remains too costly. When budget constraints are relaxed, the best assignment for a Batch-Only method eventually reduces to $b=1$, after which accuracy can no longer improve because harder queries cannot be reassigned to stronger models. By jointly optimizing model selection and batch prompting, \RoBatch avoids both limitations and achieves a better cost-accuracy frontier across a wide range of costs.

\subsection{Sensitivity \& Design Choice Analysis}
\label{sec:exp:sensitivity}
In this section, we present a series of sensitivity and design-choice analyses for \RoBatch using the Qwen3 family.
For each test configuration, we use three budget levels: the total cost of the cheapest model (Qwen3-4B), the total cost of the medium-cost model (Qwen3-14B), and their midpoint. 
The configurations are maintained as default except the test hyper-parameter or design choice.

\subsubsection{Coreset selection algorithms and coreset sizes}

\begin{figure}[t]
\centering
\includegraphics[width=0.6\columnwidth]{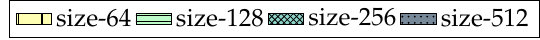}

 \begin{tabular}[h]{c}
 
        \subfigure[\GSM] {
				\includegraphics[width=0.32\columnwidth]{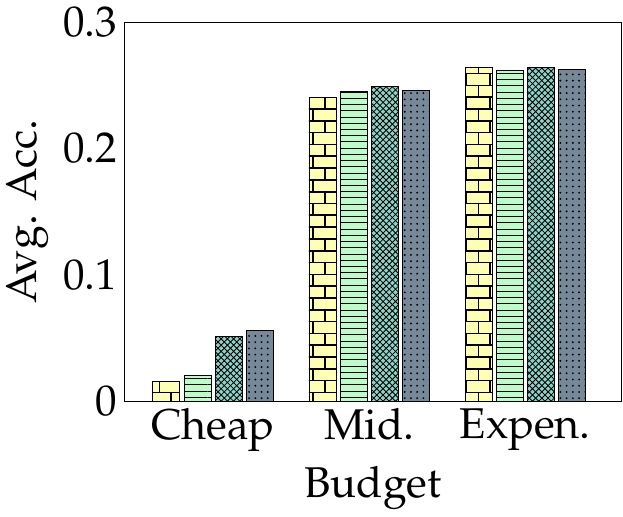}
			\label{fig:exp:ablation:gsm8k}
		} \hspace{-2.5ex}       
        \subfigure[\AGNews] {
				\includegraphics[width=0.32\columnwidth]{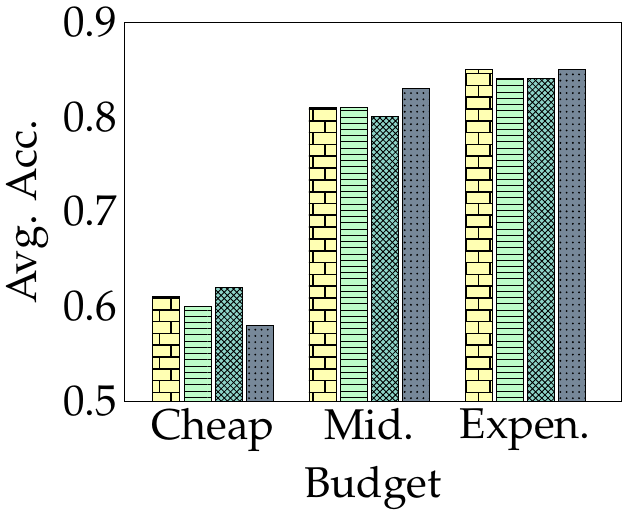}
			\label{fig:exp:coreset_size:ag_news}
		} \hspace{-2.5ex}    
        \subfigure[\IMDB] {
				\includegraphics[width=0.32\columnwidth]{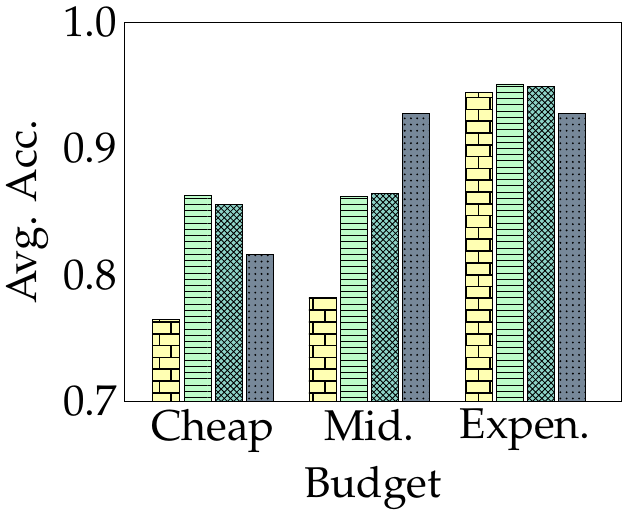}
			\label{fig:exp:ablation:imdb}
		}         

\end{tabular}

	\caption{Accuracy of Different Coreset Sizes}

    \label{fig::exp::sen::coreset::size}

\end{figure}

To evaluate the robustness of \RoBatch with respect to the construction of the coreset, we study the sensitivity of \RoBatch to the coreset selection methods and the coreset size. Specifically, we compare three widely used coreset selection algorithms, i.e., $k$-center greedy, Facility Location (FL)~\cite{DBLP:conf/interspeech/LinB09} and Herding~\cite{DBLP:conf/icml/Welling09}, and vary the coreset size $|\mathcal{Q}''|$ from $\{64, 128, 256, 512\}$.
Overall, the results in Table~\ref{tab::combined::results} and Fig.~\ref {fig::exp::sen::coreset::size} suggest that \RoBatch is robust to the coreset construction. Across different datasets and budget levels, changing the coreset selection algorithm introduces only limited perturbation to the final Accuracy, and no single method consistently dominates in all cases. For instance, FL is slightly stronger on \AGNews at the middle and expensive budgets, while $k$-center greedy and Herding remain competitive on \GSM and \IMDB, respectively, indicating that the framework is not strongly dependent on any coreset selection algorithms. In addition, increasing the coreset size from 64 to 512 results in only marginal performance differences in most settings, suggesting that even a relatively small coreset is sufficient to capture the main utility patterns needed by \RoBatch.

\begin{table}[t]
\centering
\scriptsize
\setlength{\tabcolsep}{3pt}
\caption{Accuracy of different coreset selection methods, query embedding models, and fitting methods of $\rho_k{(b)}$ under three budget levels}
\label{tab::combined::results}
\begin{tabular}{l|ccc|ccc|ccc}
\toprule
& \multicolumn{3}{c}{\AGNews} & \multicolumn{3}{c}{\GSM} & \multicolumn{3}{c}{\IMDB} \\
Method & Cheap & Mid. & Expen.
& Cheap & Mid. & Expen.  
& Cheap & Mid. & Expen. \\
\midrule
K-center   & 0.5996 & 0.8027 & 0.8438 & 0.0615 & 0.2510 & 0.2617 & 0.8633 & 0.8633 & 0.9502 \\
FL  & 0.5996 & 0.8164 & 0.8594 & 0.0508 & 0.2451 & 0.2695 & 0.8945 & 0.8936 & 0.9404 \\
Herding    & 0.6055 & 0.8037 & 0.8428 & 0.0498 & 0.2354 & 0.2627 & 0.9082 & 0.9014 & 0.9463 \\
\midrule
BGE-base        & 0.4580 & 0.8174 & 0.8555 & 0.0498 & 0.1982 & 0.2432 & 0.8682 & 0.8672 & 0.9443 \\
E5-base         & 0.6084 & 0.8037 & 0.8447 & 0.0625 & 0.2686 & 0.2705 & 0.9053 & 0.9014 & 0.9375 \\
Qwen3-0.6B  & 0.5986 & 0.8037 & 0.8457 & 0.0576 & 0.2520 & 0.2588 & 0.8652 & 0.8652 & 0.9492 \\
\midrule
Piecewise lin. & 0.6113 & 0.7998 & 0.8564 & 0.0596 & 0.2354 & 0.2656 & 0.8633 & 0.8828 & 0.9385 \\
Power-low      & 0.6035 & 0.7939 & 0.8506 & 0.0586 & 0.2354 & 0.2627 & 0.8623 & 0.8828 & 0.9395 \\
KNN lin.       & 0.6064 & 0.8115 & 0.8545 & 0.0586 & 0.2295 & 0.2627 & 0.8652 & 0.8877 & 0.9395 \\
\bottomrule
\end{tabular}

\end{table}

\subsubsection{Impact of embedding models}

To evaluate the robustness of our framework with different query embedding models, we perform a sensitivity analysis using several alternative embedding models, including E5-base-v2~\cite{DBLP:journals/corr/abs-2212-03533}, Qwen3-0.6B-embedding~\cite{DBLP:journals/corr/abs-2506-05176}, and BGE-base-en-v1.5~\cite{DBLP:journals/corr/abs-2402-03216}. As shown in Table~\ref{tab::combined::results}, the differences in the final Accuracy of \RoBatch among BGE-base, E5-base, and Qwen3-0.6B remain small across the three datasets and three budget constraints, although each embedding model shows slight advantages in specific cases. In summary, \RoBatch does not depend heavily on a particular embedding model to maintain strong overall performance.

\subsubsection{Impact of routing models and hyper-parameters}

\begin{figure}[t]
\includegraphics[width=0.8\columnwidth]{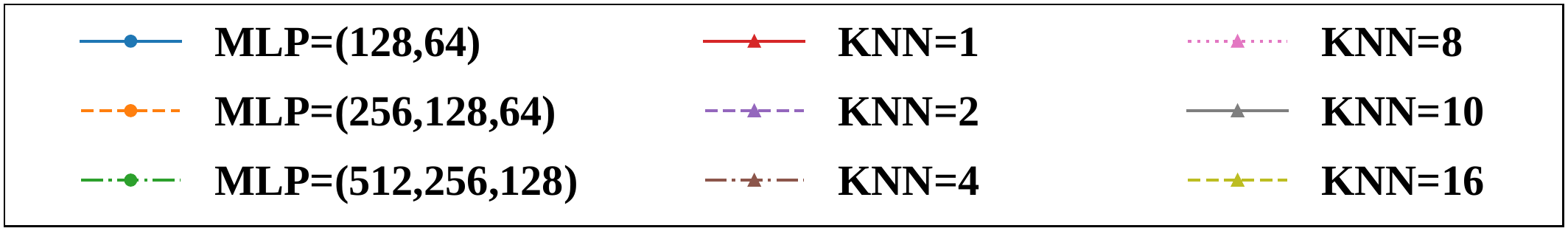}
\setlength{\tabcolsep}{0pt}

 \begin{tabular}[h]{@{}ccc@{}}
        \subfigure[\GSM] {
				\includegraphics[width=0.3\columnwidth]{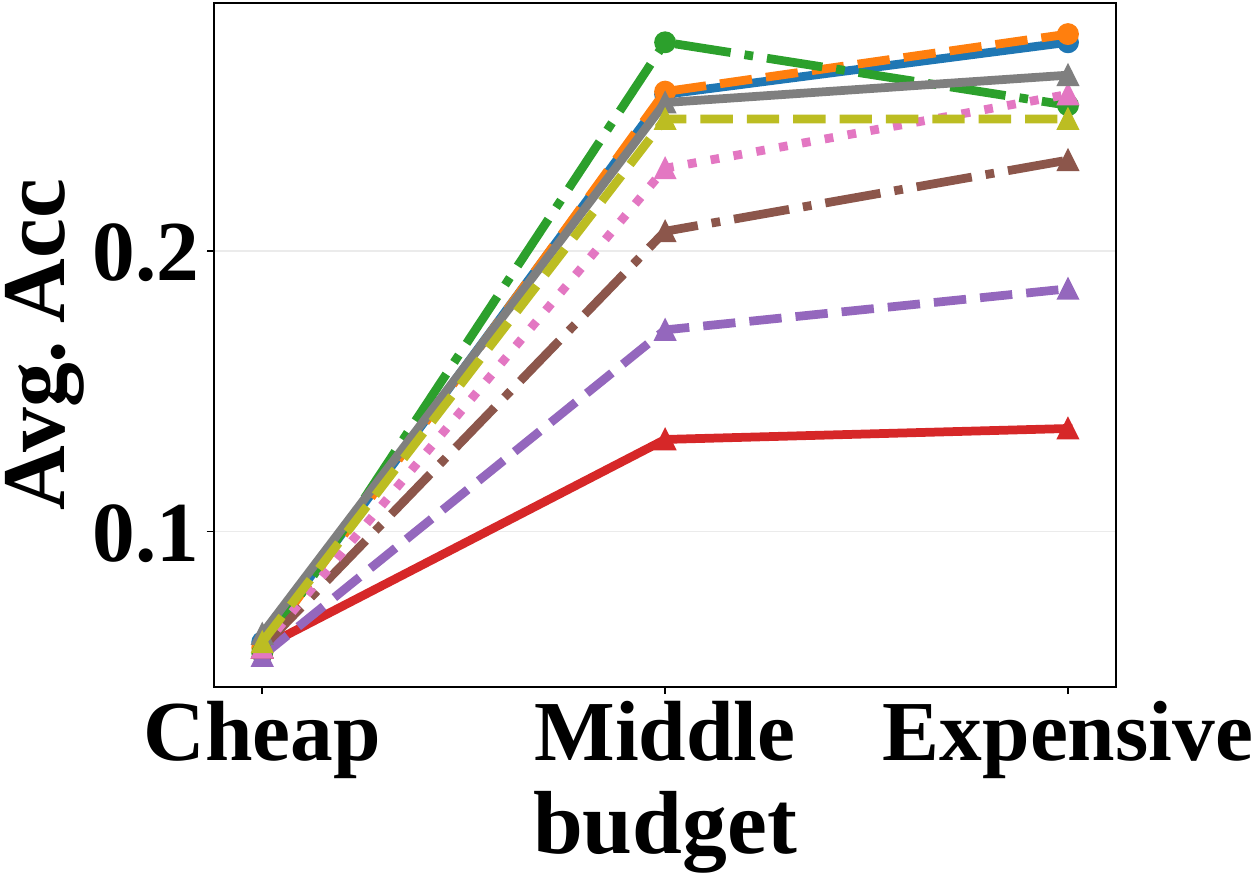}
			\label{fig:exp:sen:router:gsm8k}
		}         
        \subfigure[\AGNews] {
				\includegraphics[width=0.3\columnwidth]{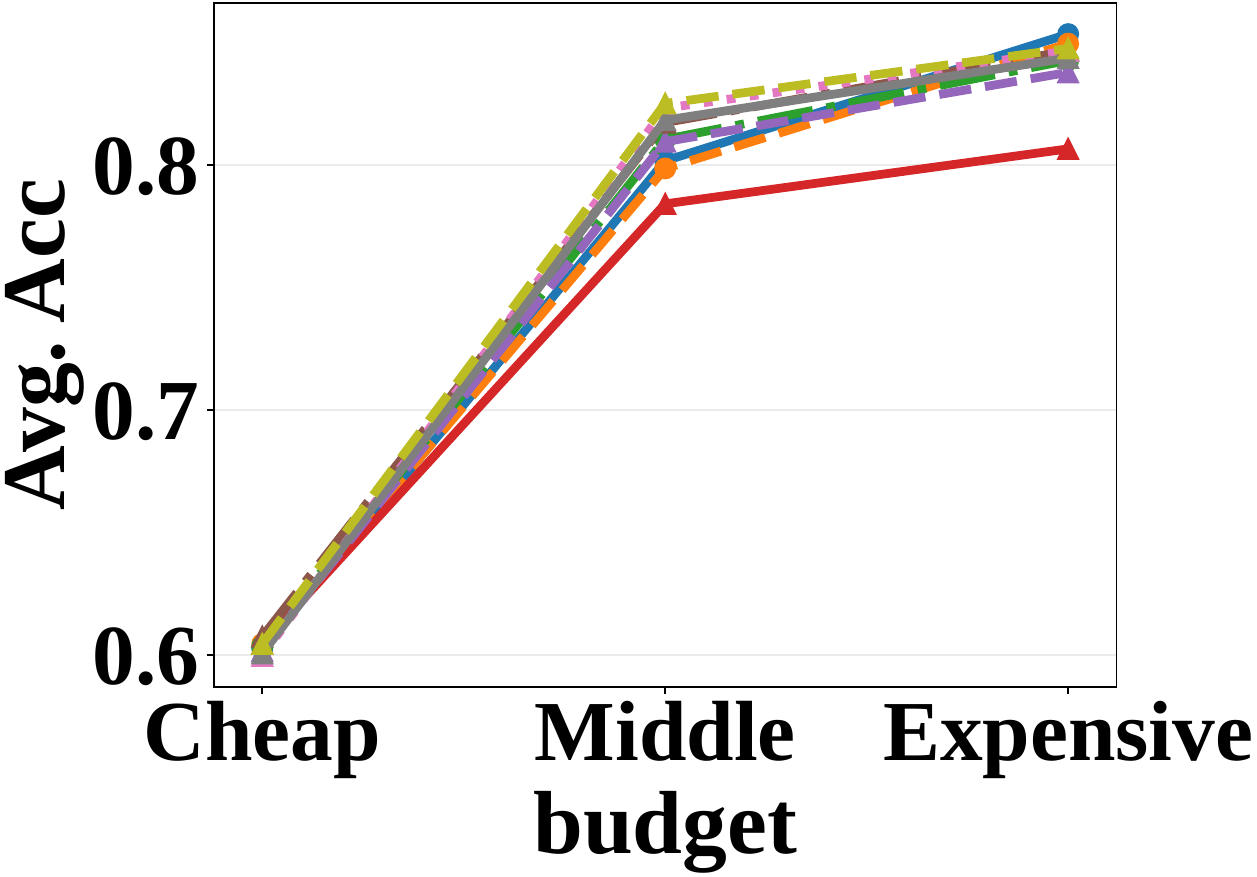}
			\label{fig:exp:sen:router:ag_news}
		}         
        \subfigure[\IMDB] {
				\includegraphics[width=0.3\columnwidth]{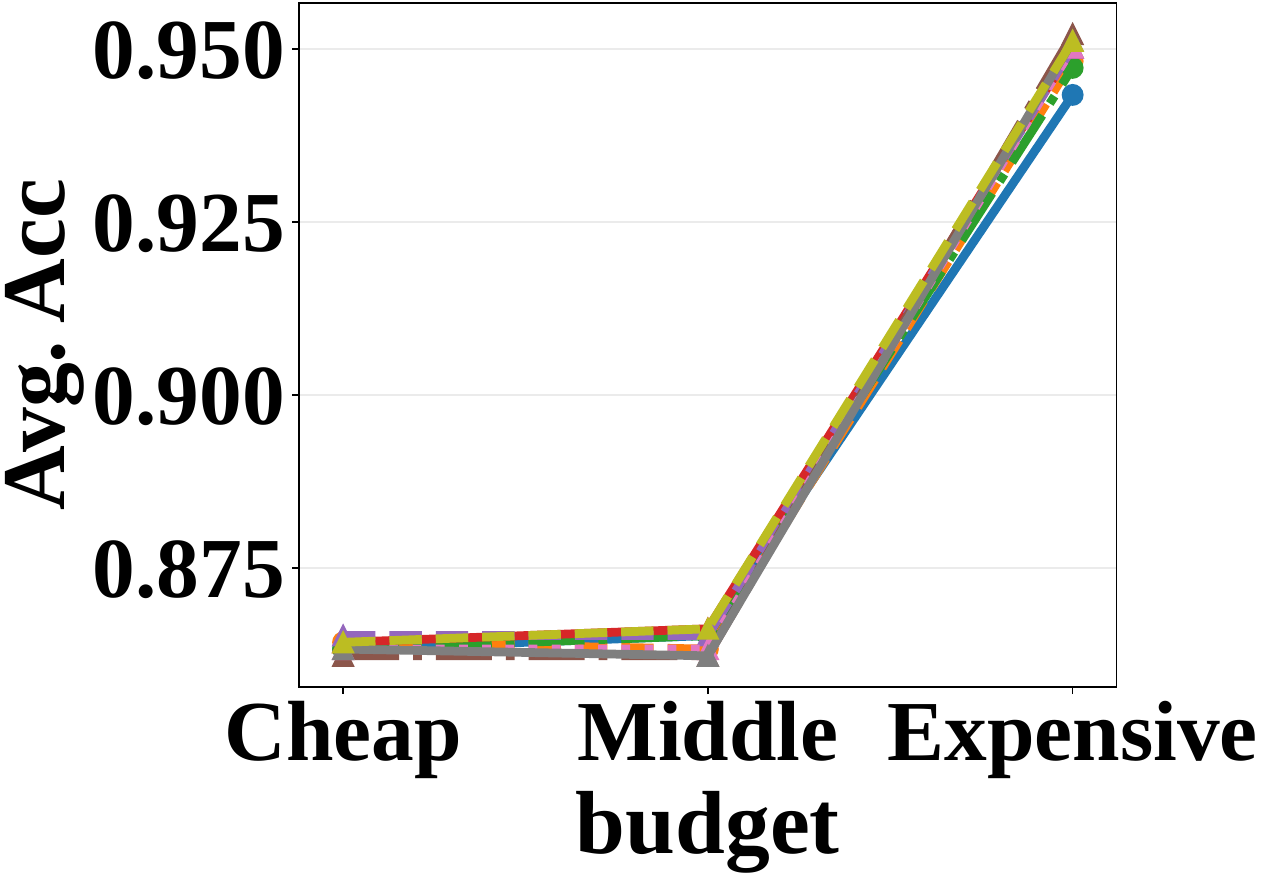}
			\label{fig:exp:sen:router:imdb}
		}         

\end{tabular}

	\caption{Sensitivity of MLP and KNN Hyper-Parameters}

    \label{fig:exp:sen:router}

\end{figure}

To further examine the robustness of \RoBatch with respect to the routing component, we study the impact of different router models and hyper-parameter settings, including MLP classifiers with different hidden-layer configurations and KNN-based classifiers with varying numbers of nearest neighbors, $k$. As shown in Fig.~\ref{fig:exp:sen:router}, \RoBatch shows limited sensitivity to the specific router models or hyperparameter choices, especially on \AGNews and \IMDB, where most MLP and KNN variants achieve very similar accuracy across the three budget levels. The effect is more evident on the more challenging \GSM workload. While different MLP model architectures yield largely comparable results, KNN is more sensitive to $k$, and typically performs better with a moderate $k$; in contrast, the extreme setting $k=1$ leads to clearly inferior accuracy.

\subsubsection{Impact of different scaling functions}

We investigate three representative ways of fitting the utility scaling function $\rho_k(b)$ to evaluate their impact on overall Accuracy under different budget constraints. 
We compare the default piecewise linear interpolation, as shown in  Eq.~\eqref{eq:linear_interp} with two alternative fitting methods: (1) Power-law function fitting that models the decay of utilities as $\rho_k(b) = 1 - \alpha(b - 1)^\beta$, and fits $\alpha, \beta$ by nonlinear least squares method on the coreset $\mathcal{Q}''$. (2) KNN linear interpolation retrieves the nearest neighbors from $\mathcal{Q}''$ for each test query and utilizes the average utility of these neighbors at various batch sizes as query-specific interpolation points. 
The results in Table~\ref{tab::combined::results} show that the three fitting methods yield highly consistent performance, demonstrating the robustness of \RoBatch to decay function formulations. Although different methods exhibit relative advantages under different scenarios, the differences in final Accuracy remain within 2\%. Overall, across three different workloads and three budget constraints, the convergence of all functions confirms that default piecewise linear interpolation is sufficient for robust practical deployment.
The overlapping cost-accuracy curves indicate that performance gains stem primarily from the joint optimization of model selection and batching rather than a specific local design.







\subsection{Scalability \& System Overhead}
\label{sec:exp:scalability}

\subsubsection{Scalability comparison with baselines}
\begin{figure}[t]
\includegraphics[width=0.7\columnwidth]{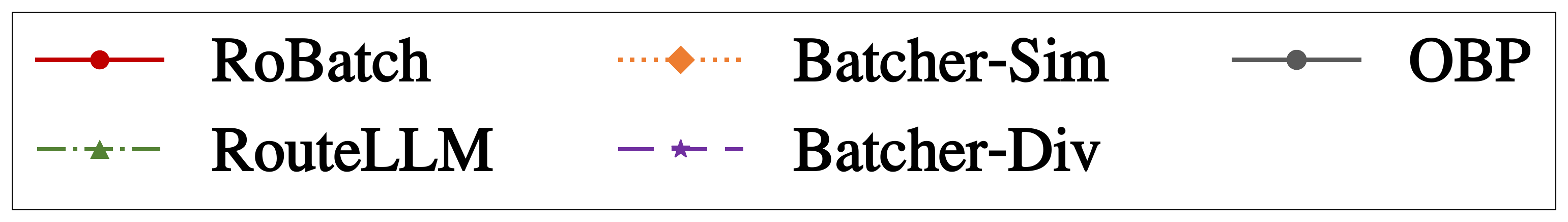}
\setlength{\tabcolsep}{0pt}

 \begin{tabular}[h]{@{}ccc@{}}
        \subfigure[\MMLU] {
				\includegraphics[width=0.32\columnwidth]{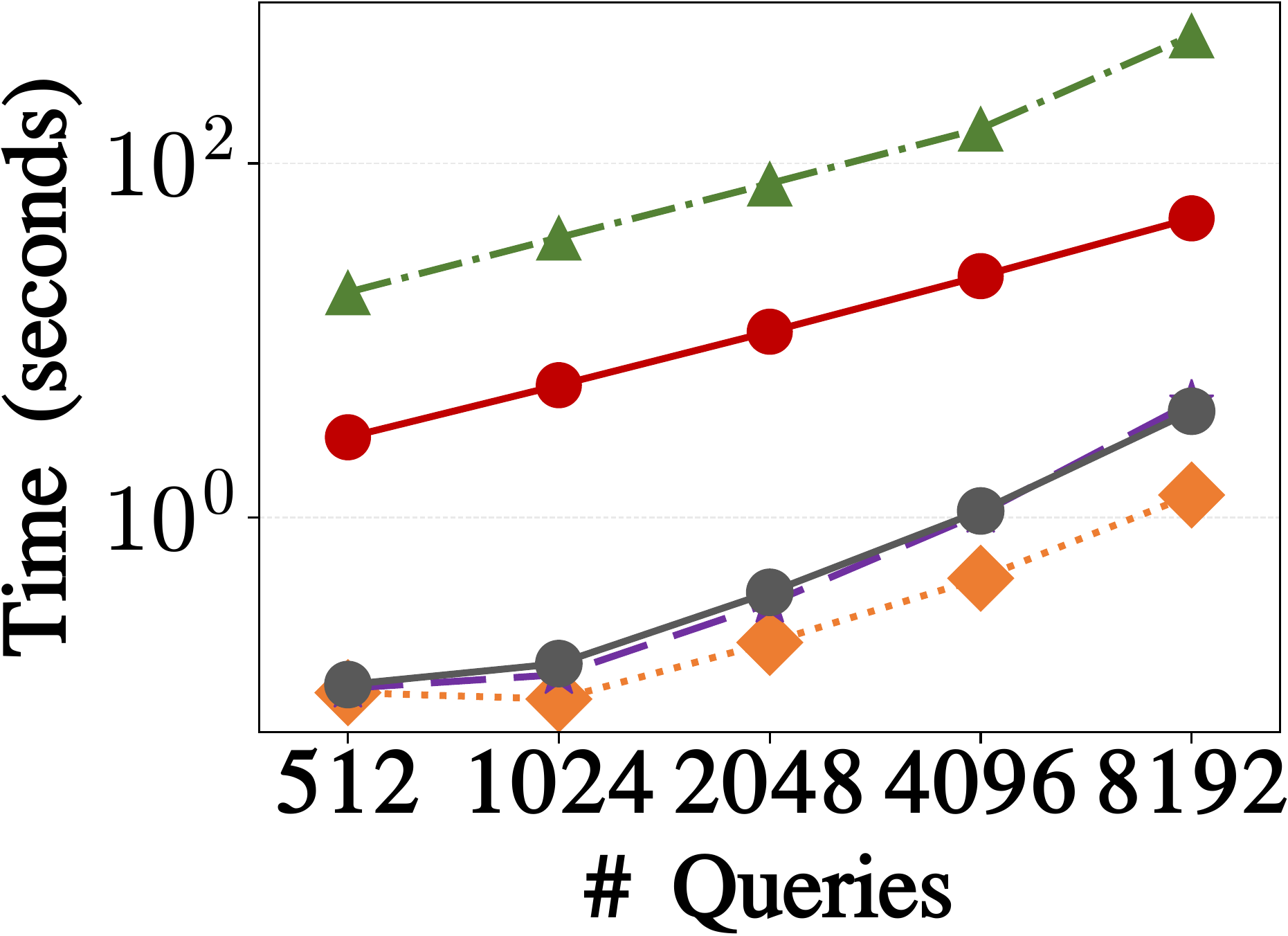}
			\label{fig:exp:scalability:method:gsm8k}
		} \hspace{-2ex}         
        \subfigure[\AGNews] {
				\includegraphics[width=0.32\columnwidth]{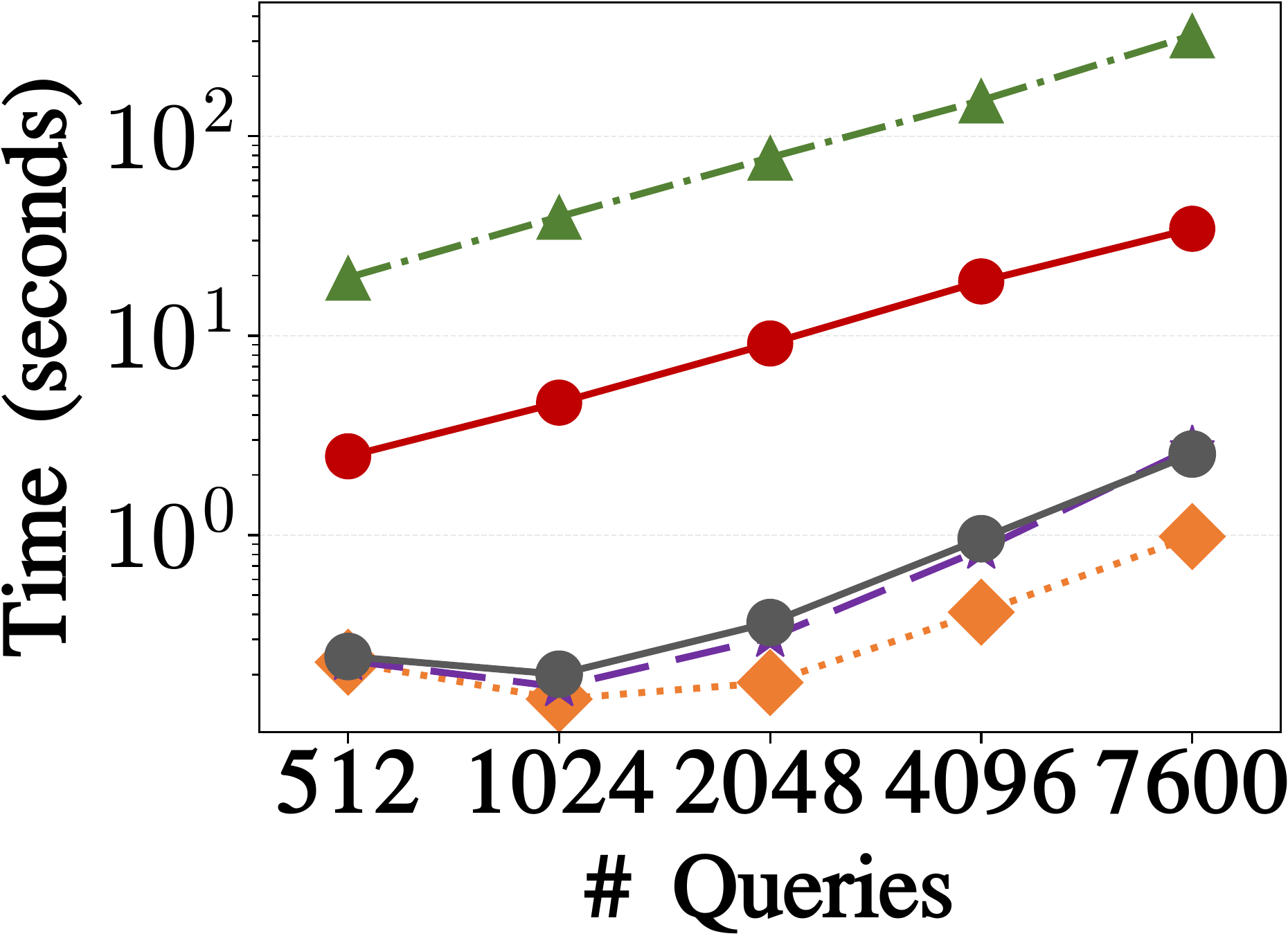}
			\label{fig:exp:scalability:method:ag_news}
		} \hspace{-2ex}        
        \subfigure[\IMDB] {
				\includegraphics[width=0.32\columnwidth]{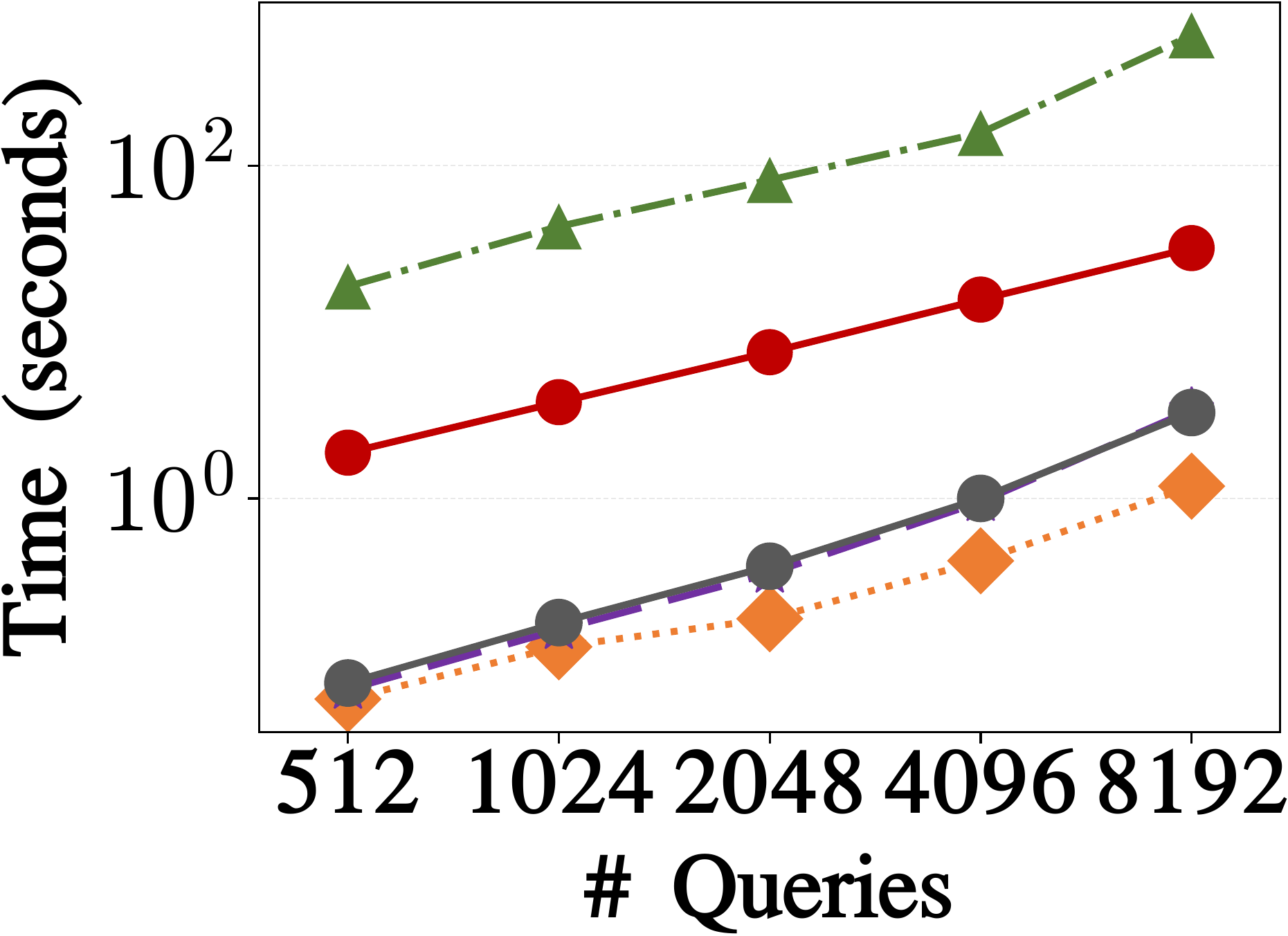}
			\label{fig:exp:scalability:method:imdb}
		}         

\end{tabular}

	\caption{Comparison of Latency and Scalability}

    \label{fig:exp:scalability:method}

\end{figure}

We compare how the serving overhead of different methods varies with workload size, which is measured as the elapsed time from receiving the test queries to issuing the practical LLM requests.
Fig.~\ref{fig:exp:scalability:method} shows this elapsed time of \RoBatch and three baselines, as the number of queries grows. 
\FrugalGPT is excluded from this comparison because its LLM cascading design makes LLM API latency and the scheduling time intertwined. 
Compared with \BATCHER baselines, \RoBatch incurs higher overhead because it performs additional router prediction and budget-aware greedy scheduling. 
As the workload size doubles,
the serving overhead of \RoBatch grows approximately linearly, rising from 2.48s to 34.35s on \AGNews, from 1.89s to 31.97s on \IMDB, and from 2.84s to 48.95s on \MMLU.
\RouteLLM requires computing similarities against the entire training dataset and then solving a weighted optimization problem for each test query, which results in higher latency.
In contrast, both the \BATCHER and \OBP begin with a computationally router prediction; \BATCHER then applies low-complexity $k$-means clustering, while \OBP maintains low latency through efficient clustering and lightweight approximate splitting instead of expensive global optimization.
In summary, \RoBatch exhibits a predictable scaling behavior as the workload size increases.
%

\subsubsection{Latency breakdown of \RoBatch}
\begin{figure}[t]
\includegraphics[width=0.8\columnwidth]{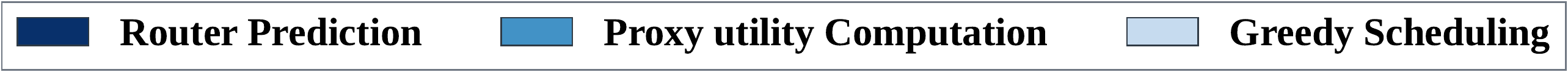}
\setlength{\tabcolsep}{0pt}

 \begin{tabular}[h]{@{}ccc@{}}
        \subfigure[\MMLU] {
				\includegraphics[width=0.32\columnwidth]{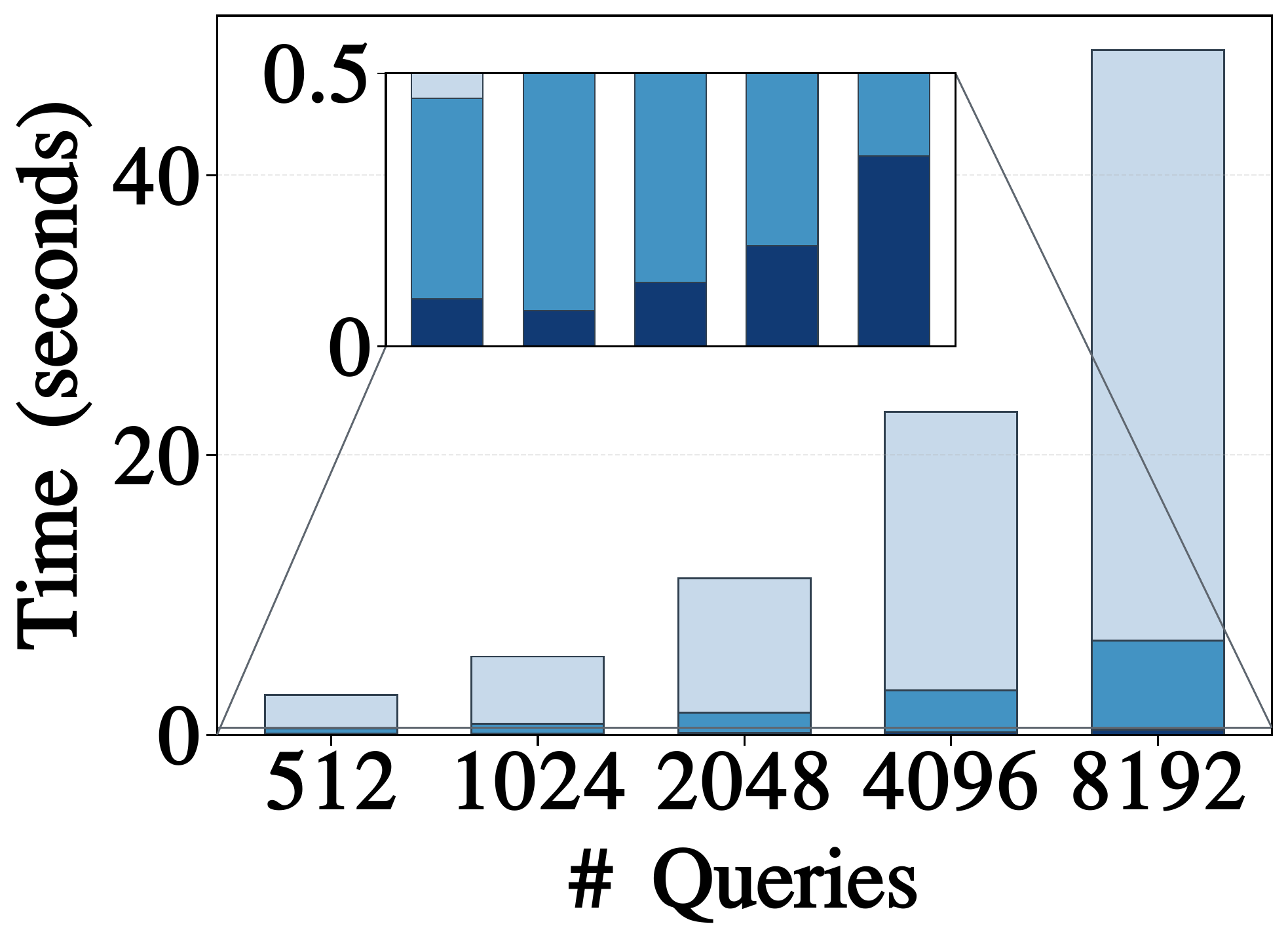}
			\label{fig:exp:scalability:size:gsm8k}
		}
        \hspace{-2ex}
        \subfigure[\AGNews] {
				\includegraphics[width=0.32\columnwidth]{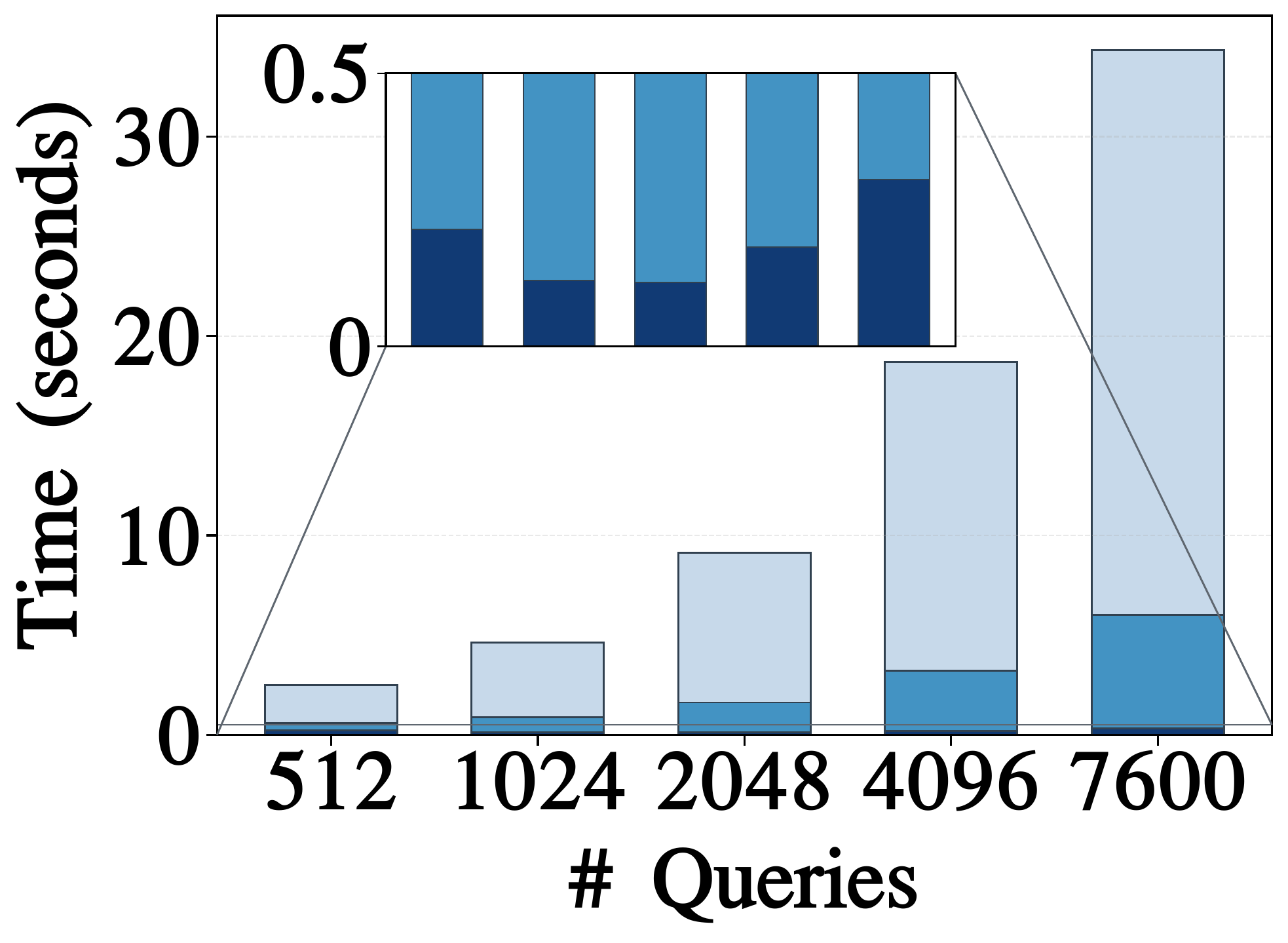}
			\label{fig:exp:scalability:size:ag_news}
		}
        \hspace{-2ex}
        \subfigure[\IMDB] {
				\includegraphics[width=0.32\columnwidth]{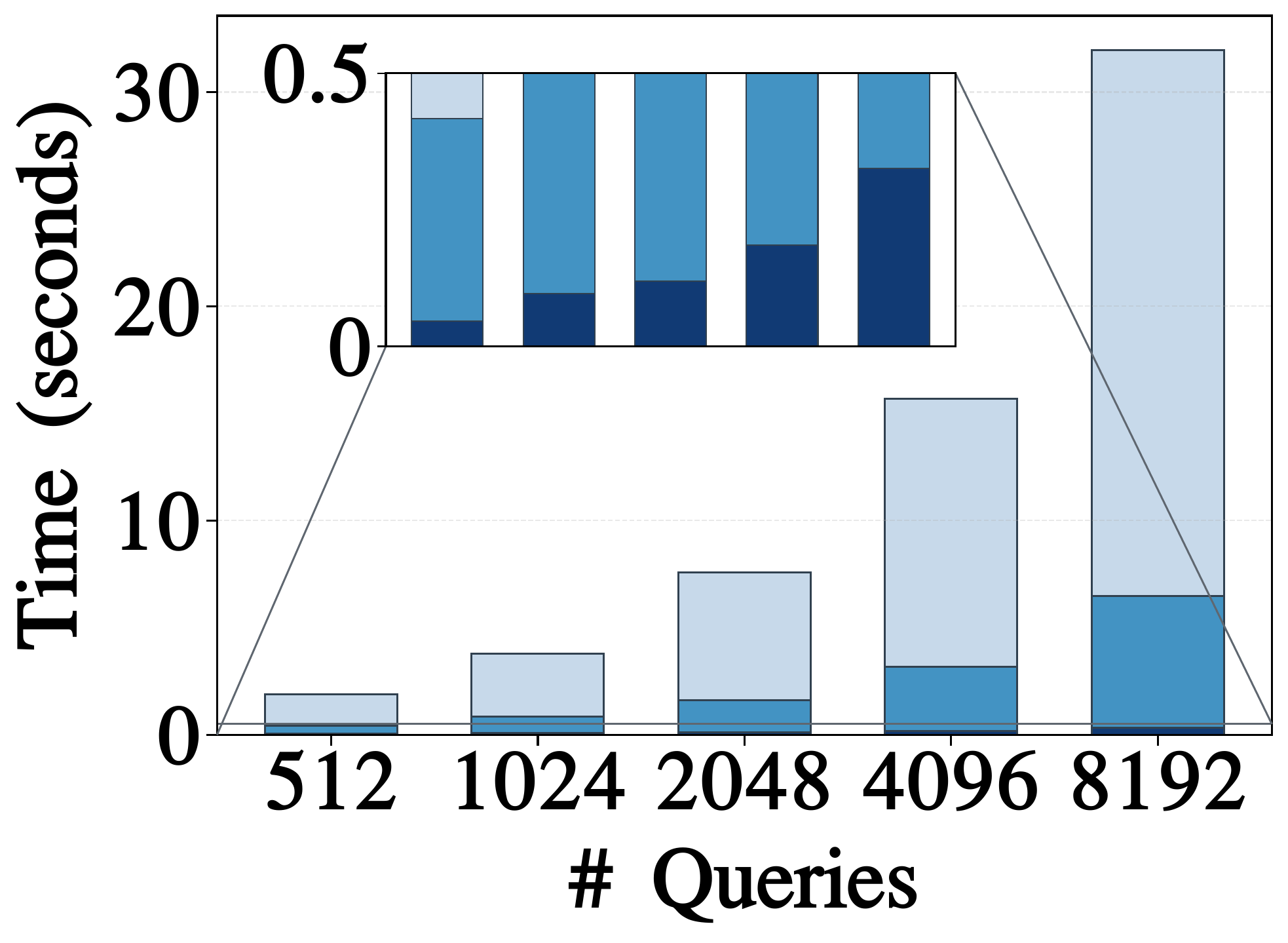}
			\label{fig:exp:scalability:size:imdb}
		}         

\end{tabular}

	\caption{Latency Breakdown of \RoBatch}

    \label{fig:exp:scalability:size}

\end{figure}

Fig.~\ref{fig:exp:scalability:size} further decomposes the serving overhead of \RoBatch into three parts: router prediction, proxy utility computation, and greedy scheduling. The profiling results reveal a clear and stable pattern: greedy scheduling dominates the total overhead, proxy utility computation is the second largest component, and router prediction is negligible. Concretely, greedy scheduling accounts for  76.0\%-82.8\% of the total time on \AGNews, 77.9\%-79.7\% on \IMDB, and 84.0\%-86.3\% on \MMLU, while proxy utility computation accounts for  15.4\%-16.6\%, 19.1\%-19.7\%, and 12.9\%-13.1\%, respectively. This breakdown is consistent with the complexity analysis in \cref{sec:routing}, where the $\mathcal{O}(|\mathcal{Q}|Kd)$ router prediction is lightweight, and the $\mathcal{O}(|\mathcal{Q}|T\log |\mathcal{Q}|)$ greedy scheduling term is the dominant factor in practice.

\section{Related Work}
 \label{sec:rw}

\stitle{LLM Routing and Model Selection.}
Routing queries across a diverse pool of LLMs has emerged as an effective strategy to optimize the trade-off between task accuracy and monetary cost~\cite{chen2026workload}. Existing research focuses on developing LLM performance estimators that  dynamically predict the suitability of candidate models for individual queries. \FrugalGPT~\cite{DBLP:journals/tmlr/ChenZ024} introduced the concept of LLM cascades, which reduce costs by sequentially invoking models of increasing capacity based on uncertainty calibration. Hybrid LLM~\cite{DBLP:conf/iclr/DingM0SMRLA24} extends the routing paradigm to edge-cloud collaborative environments, employing a router to assess query difficulty and process simple queries at the edge while directing complex tasks to cloud-based models. To improve routing reliability, AutoMix~\cite{DBLP:conf/nips/AggarwalMAPMZGR24} introduces a self-verification mechanism where a smaller model evaluates the confidence of its own response to decide whether to escalate to a more capable model. Furthermore, \RouteLLM~\cite{DBLP:conf/iclr/OngAWC0GKS25} explores learning routing strategies from preference data by predicting the marginal benefit of invoking a stronger model to close the performance gap between weak and strong models. These frameworks collectively establish paradigms for cost-effective model routing grounded in difficulty prediction, confidence calibration, and preference-driven learning.

\stitle{Batch Prompting.}
Batch prompting~\cite{DBLP:conf/emnlp/ChengK023, DBLP:conf/iclr/LinDDA24} serves as a key optimization for improving the inference efficiency of LLM APIs by aggregating multiple independent queries into a single request. 
Early studies demonstrate that such aggregation effectively amortizes the fixed overhead of system prompts and reduces the total number of API calls for various reasoning tasks.
To further explore this technique, \BATCHER~\cite{DBLP:conf/icde/FanHFC00024} provides a comprehensive design space exploration specifically for In-Context Learning for tasks like Entity Resolution. This work introduces contrasting grouping principles: \BATCHERSIM clusters semantically similar queries, whereas \BATCHERDIV prioritizes query diversity to enhance performance. Building on these heuristics, Optimized Batch Prompting (\OBP)~\cite{DBLP:journals/pvldb/JiWLXZ25} shifts from fixed-size batching to an adaptive query batching  framework. By leveraging clustering and dynamic refinement, \OBP balances semantic affinity, context length, and accuracy to minimize costs.
Overall, the existing literature establishes a solid foundation for enhancing LLM efficiency through batch prompting, with efforts primarily focused on refining query grouping heuristics and adaptive clustering to balance the economic benefits of system prompt amortization against task-specific performance.
\section{Conclusion}
\label{sec:conclusion}

LLM routing and batch prompting are two independent techniques for LLM inference optimization. 
In this paper, we study cost-effective LLM routing with batch prompting that simultaneously determines which model should serve each query and how to group queries into batches.
We formulate the budget-constrained route with batching problem as a combinatorial optimization and prove that the problem is NP-hard. 
To address this hard problem, we propose \RoBatch, a unified framework that integrates batch-aware utility modeling with budget-aware greedy scheduling. \RoBatch first builds a batch-aware proxy utility model that estimates how utility changes across models and batch sizes, and then uses this model to guide routing decisions under a given cost budget.
Extensive experiments on six benchmarks show that \RoBatch consistently achieves a better cost–performance frontier than routing-only and batching-only baselines, while incurring only modest scheduling overhead in practice.

\balance
\bibliographystyle{ACM-Reference-Format}
\bibliography{ref}

\end{document}